# The genomic architecture and evolutionary fates of supergenes

Juanita Gutiérrez-Valencia[1,§], William Hughes[1,§], Emma L. Berdan[1], Tanja Slotte[1*]

[1]Department of Ecology, Environment and Plant Sciences, Science for Life Laboratory, Stockholm University

[*]Corresponding author:
Tanja Slotte
Department of Ecology, Environment and Plant Sciences
Stockholm University
SE-106 91 Stockholm, Sweden

Email: tanja.slotte@su.se
Phone: +46-76-125 21 09

[§]These authors contributed equally

Word count for main text, excluding references: 8,536

## Abstract

Supergenes are genomic regions containing sets of tightly linked loci that control multi-trait phenotypic polymorphisms under balancing selection. Recent advances in genomics have uncovered significant variation in both the genomic architecture as well as the mode of origin of supergenes across diverse organismal systems. Although the role of genomic architecture for the origin of supergenes has been much discussed, differences in the genomic architecture also subsequently affect the evolutionary trajectory of supergenes and the rate of degeneration of supergene haplotypes. In this review, we synthesize recent genomic work and historical models of supergene evolution, highlighting how the genomic architecture of supergenes affects their evolutionary fate. We discuss how recent findings on classic supergenes involved in governing ant colony social form, mimicry in butterflies, and heterostyly in flowering plants relate to theoretical expectations. Furthermore, we use forward simulations to demonstrate that differences in genomic architecture affect the degeneration of supergenes. Finally, we discuss implications of the evolution of supergene haplotypes for the long-term fate of balanced polymorphisms governed by supergenes.

## Keywords

Balancing selection, degeneration, hemizygosity, inversion, recombination suppression, structural variation

## Significance Statement

Supergenes are sets of tightly linked loci that together control complex (i.e. multi-trait) balanced polymorphisms. Many iconic polymorphisms, such as the pin and thrum floral morphs in plants, and polymorphic warning coloration in butterflies, are controlled by supergenes. While the evolution of supergenes has long interested evolutionary biologists, little was known about their genomic architecture, but this is fast changing following the genomic revolution. Here, we review recent genomic studies on classic supergenes and discuss how different genomic architectures shape the evolutionary trajectories of supergene haplotypes. Additionally, we use simulations to explore how the maintenance of balanced polymorphisms might affect the evolution of supergene haplotypes.

## Introduction

The maintenance of favorable combinations of traits is essential for the evolution of sex determination, mating systems, local adaptation, and speciation (reviewed by Thompson and Jiggins 2014), but maintenance of these combinations in the face of recombination is problematic. Supergenes present one solution to this problem; they are tightly linked sets of loci, which control complex (i.e. multi-trait) adaptive phenotypic polymorphisms under balancing selection (Thompson and Jiggins 2014). The genomic architecture of a supergene refers to the size of the non-recombining region, the gene content or density of selected sites, and type (if any) of structural variation that it harbors. Although the supergene concept has been prevalent in the literature since the Modern Synthesis, recent advances in genomics have brought the field into a new era (reviewed by Schwander et al. 2014). In particular, genomic methods have allowed us to elucidate the genomic architecture of mimicry in insects, plant self-incompatibility, and colony organization in social organisms (reviewed by Schwander et al. 2014; Charlesworth 2016a) (Box 1). These studies have found that the genomic architecture underlying complex balanced polymorphisms sometimes differs markedly from expectations under classic supergene models (reviewed by Booker et al. 2015; Charlesworth 2016a). They have also documented substantial variation among systems in the genomic architecture and mode of origin of supergenes (Table 1). While the role of genomic architecture in the origin of supergenes has been much discussed (reviewed in Schwander et al. 2014; Charlesworth and Charlesworth 2010; Thompson and Jiggins 2014; Charlesworth 2016a), recent genomic studies have highlighted its impact on the continued evolution of supergenes and their evolutionary fate (Table 1). These results allow us to begin to connect genomic architecture to long-term consequences for supergene evolution and maintenance.

In this review, we first provide a brief overview of the history of the supergene concept and models of supergene origins. We compare and contrast different genomic architectures and discuss theoretical predictions for their impact on the evolutionary fate of supergenes. We then discuss recent findings on classic supergenes that govern balanced polymorphisms, highlighting the genomic architecture and its consequences for supergenes that control ant social colony form, mimicry in butterflies, and heterostyly in plants. We compare these results with new forward simulations explicitly examining the role of genomic architecture in the degeneration of supergenes. Finally, we discuss implications of supergene genomic architecture for the evolution of supergene haplotypes and the long-term maintenance of balanced polymorphisms.

## Supergenes – history and definitions

The supergene concept arose early during the Modern Synthesis, to explain the apparent conundrum of complex balanced polymorphisms that were inherited as if they were governed by a single Mendelian factor. Fisher, looking to explain the genomic architecture of Batesian mimicry in butterflies (Figure 1), suggested that the basis of this complex polymorphism could be a set of genes kept in linkage disequilibrium by suppressed recombination (Fisher 1930). Likewise, Ernst (1936), who studied the genetic basis of heterostyly in primroses (Figure 1), argued that a system of three closely coupled yet distinct genes determined the discrete pin (L-morph) and thrum (S-morph) floral types in heterostylous *Primula* species. The discovery of widespread inversion polymorphism in *Drosophila* subsequently provided a plausible mechanism for recombination suppression (Sturtevant and Beadle 1936; Dobzhansky and Sturtevant 1938; Wright and Dobzhansky 1946; Dobzhansky and Epling 1948). The term *supergene* was coined by Darlington and Mather (1949), although the concept originated earlier. In modern terms, the classic multi-gene supergene model posits that a supergene is a genomic region that contains closely linked loci protected from recombination, and therefore sets of alleles at these loci (i.e. haplotypes) are consistently inherited together (reviewed by Schwander et al. 2014; Charlesworth 2016a).

Recent genomic studies have shown that some polymorphisms thought to be governed by classic multi-gene supergenes involve divergent alleles of major effect at single genes (Kunte et al. 2014; Nishikawa et al. 2015). However, even in cases where a single gene is implicated, changes at multiple *cis*-regulatory elements may have been involved (Thompson and Jiggins 2014). To include both classic supergenes involving multiple closely linked genes and cases that involve changes at multiple tightly linked sites within a single gene, Thompson and Jiggins (2014) proposed an updated definition of supergenes as "a genetic architecture involving multiple linked functional genetic elements that allows switching between discrete, complex phenotypes maintained in a stable local polymorphism".

A benefit of this broader definition is that it allows a variety of potential genomic architectures of supergenes to be considered. As such, we use the definition of Thompson and Jiggins (2014) to discuss how different genomic architectures responsible for distinct polymorphisms may modify expectations regarding evolutionary genetic patterns. We will refrain from discussing empirical studies of sex chromosomes in depth as there are several recent reviews on this topic (Charlesworth 2016b; Wright et al. 2016; Palmer et al. 2019; Vicoso et al. 2019; Furman et al. 2020), although sex chromosomes have long been viewed as one form of supergene (see e.g. recent review by Charlesworth 2016a). Finally, we will as far as possible restrict our discussion to supergenes involved in governing balanced polymorphisms within local populations, although there are interesting parallels with genomic architectures of local adaptation (Kirkpatrick and Barton 2006).

## How do supergenes form? Classic models and recent extensions

One of the most challenging questions in supergene evolution is that of their origin. Supergene formation requires the establishment of mutations at a minimum of two loci, that together result in co-segregation of beneficial trait combinations. Our current understanding of how such tightly linked multi-locus polymorphisms might be established is mostly based on theoretical models of the evolution of sex chromosomes and supergenes that govern Batesian mimicry (Charlesworth and Charlesworth 1976, 1978). A common feature of these models is that they involve sequential establishment of "modifier" mutations that are beneficial only in genotypes carrying alleles at specific other loci that are already polymorphic due to balancing selection (reviewed by Charlesworth 2016a). The sequential establishment of linked mutations to form a region harboring a multilocus, multiallele polymorphism under balancing selection is thus central to classic models of supergene and sex chromosome evolution (see Figure 9.4 in Charlesworth and Charlesworth 2010, and Figures 1 and 2 in Charlesworth 2016a). Additional modifier mutations with antagonistic effects in different morphs or sexes (e.g. sexually antagonistic alleles) are more likely to invade the population if they are in linkage disequilibrium with the initial polymorphism. Such antagonistic polymorphisms can lead to further selection for suppressed recombination (Rice 1987, Jordan and Charlesworth 2012). Stepwise cessation of recombination, consistent with this model, has indeed been observed in independently derived sex chromosome systems (Lahn and Page 1999). Nijhout (1994) proposed a further refinement of the molecular underpinnings of the classic supergene model. He suggested that genetic changes in a limited number of linked regulatory loci could, through their effects on structural genes, produce major-effect changes. Accordingly, stepwise changes at just a few already linked regulatory loci could be sufficient to form a supergene, without a requirement for subsequent changes in gene order or selection for closer linkage (Nijhout, 1994).

Alternative models for the origin of supergenes have also been considered. For instance, structural rearrangements increasing linkage disequilibrium between previously unlinked loci could contribute to supergene formation (Turner 1967). However, in scenarios with strong balancing selection for specific trait combinations in a local population, very strong selection would be required for such rearrangements to be favored (e.g. the case of Batesian mimicry supergenes; Charlesworth and Charlesworth 1976). There may be exceptions, in the case of local adaptation, where migration leads to the introduction of immigrant alleles into the population, an inversion that captures locally adapted alleles is expected to spread (Kirkpatrick and Barton 2006; Charlesworth and Barton 2018). This process can lead to fixation of alternative inversions in different populations, but it can also lead to a stable polymorphism within a local population, for instance if the inversion captures recessive deleterious mutations as well as locally adaptive ones (Kirkpatrick and Barton 2006). A second alternative model for supergene formation was proposed by Yeaman (2013) who used simulations to show that clustered gene architectures can arise through gene transposition, under similar scenarios of local adaptation and migration (Yeaman 2013). Finally, recent findings in *Heliconius numata* and *Anopheles* suggest that there might be another important yet poorly characterized mechanism by which supergenes form, namely through the introgression of an inverted arrangement from another species (Fontaine et al. 2015; Jay et al. 2018). The idea here is that the introgression of an inverted arrangement from a separate lineage, that has already accumulated adaptive differences relative to the standard arrangement, might facilitate establishment of a supergene polymorphism (Jay et al. 2018). As we learn more about the diverse genomic architectures and evolutionary histories of supergenes, it is likely that further theoretical and simulation-based studies will be needed to investigate a broader range of possible scenarios of supergene origins.

## Molecular mechanisms of recombination suppression at supergenes

The mechanism of recombination suppression in a supergene is critical both for its origin and for its evolutionary fate. In order for supergenes to evolve, associations of beneficial or coadapted alleles need to be protected from being shuffled by recombination. We might therefore expect supergene formation to occur more readily when there is already tight linkage among loci, for instance in genomic regions with low recombination rates, such as centromeric or pericentromeric regions, or when nearby genes are involved in supergene formation (reviewed in Schwander et al. 2014). Alternatively, suppression of recombination may be mediated by structural variation. There has long been a focus on inversions as a likely mechanism for reduced recombination at supergenes, but it is now becoming increasingly clear that other types of structural variants also contribute to supergene formation (see Table 1). For instance, large hemizygous regions, genomic fragments that are only present in one of the supergene haplotypes, have been implicated as the genomic architecture of supergenes. Examples of hemizygous supergenes are found both in plants (heterostyly in *Primula*; Li et al. 2016; see section on genomic architectures of supergenes) and animals (pea aphid male wing dimorphism; Li et al. 2020; cryptic coloration morphs in *Timema* stick insects; Villoutreix et al. 2020) (Table 1). Such hemizygous supergenes may arise as a result of either duplication of genomic regions, deletion of a segment from one haplotype, or introgression. In addition to large structural variants, smaller-scale structural variation as well as epigenetic effects can also contribute to suppressed recombination (reviewed by Schwander et al. 2014).

Structural variation at supergenes strongly influences recombination. In inversions, single crossovers in heterokaryotypes lead to unbalanced gametes and thus only gene conversion and double crossovers contribute to recombination between arrangements (Crown et al. 2018). However, recombination in inversion homokaryotypes proceeds normally. In contrast, indel heterokaryotypes will form unpaired DNA loops preventing crossovers between arrangements (Poorman et al. 1981). Furthermore, recombination will only occur in one of the two homokaryotypes. Thus, assuming a similar frequency of heterozygotes, an indel will experience a greater drop in recombination than an inversion. These differences will affect accumulation of mutations at the supergene (see section on Degeneration of supergene haplotypes and Box 2).

One complication when inferring mechanisms of recombination suppression is that structural differences between non-recombining haplotypes can be either a cause or a consequence of supergene formation, as low recombination limits the impact of purifying selection, which might allow further accumulation of insertions and deletions in the region. To distinguish between these possibilities, analyses of alternative haplotypes in a comparative framework can be helpful. For instance, a comparative study showed that in *Neurospora tetrasperma* large inversions present in the *mat* chromosome are a consequence instead of a cause of suppressed recombination (Sun et al. 2017; and see additional examples in Vicoso et al. 2013; Yan et al. 2020). Likewise, population-level studies of young supergenes (see e.g. Table 1) can contribute to a more refined characterization of the early stages of supergene evolution, as has been the case for young sex chromosomes (for instance in guppies: Bergero et al. 2019; Fraser et al. 2020, see also reviews by Charlesworth 2019 and Furman et al. 2020).

## Degeneration of supergene haplotypes – theory and expectations

The low recombination necessary for the evolution of supergenes also causes them to degenerate over time by accumulating deleterious mutations, repetitive elements, and deletions (Table 1, Signatures of degeneration). But why and when do we expect this to occur and how does this impact the evolutionary trajectory of a supergene? The evolution of non-recombining regions has attracted considerable attention from evolutionary geneticists (Barton 1995; Kaiser and Charlesworth 2009; Good et al. 2014). Much of this work has centered on understanding the processes involved in the repeated decay of independently derived non-recombining Y and W sex chromosomes (reviewed in Charlesworth and Charlesworth 2000; Bachtrog 2013), and these insights can guide our expectations regarding the evolution of supergenes. Three key factors govern the extent of mutation accumulation in supergenes: (1) Recombination in the supergene region and specifically recombination between arrangements (e.g. through double crossovers or gene conversion in the case of inversions, Navarro et al. 1997), (2) The frequency of supergene haplotypes, and (3) The degree of dominance of mutations in the supergene.

Reduced recombination is a double-edged sword: it can facilitate adaptive evolutionary processes, but may also speed up the accumulation of deleterious alleles. This is because suppression of recombination reduces the efficacy of selection leading to the degeneration of non-recombining haplotypes via several processes, including Muller’s ratchet, genetic hitchhiking, and Hill-Robertson interference (reviewed by Charlesworth and Charlesworth 2000). Below, we will briefly outline how evolutionary parameters related to the genomic architecture of supergenes (e.g. number of selected sites, and strength of selection on them) and to the study system in general (e.g. mutation rate) affect the rate of degeneration.

Muller’s ratchet (Muller 1964) describes the stochastic loss of the least loaded class of chromosomes (those with the fewest deleterious mutations) in a finite population of non-recombining chromosomes. Successive losses of the least-loaded haplotypes are irreversible in the absence of recombination, and the process results in gradual accumulation of deleterious mutations. This process should be especially important in small populations and for non-recombining regions with many sites under selection, especially with high mutation rates (Bachtrog 2008).

While Muller's ratchet only operates in non-recombining regions, several other processes also contribute to degradation in regions of low recombination. For example, several forms of selection at linked sites are also thought to be key. Under background selection, purifying selection against deleterious mutations also removes linked neutral polymorphisms (Charlesworth et al. 1993). The resulting local reduction of $N_e$ may further lead to an elevated rate of fixation of mildly deleterious mutations as well as a reduction in the rate of fixation of weakly advantageous mutations (reviewed by Charlesworth and Charlesworth 2000). Under genetic hitchhiking, positively selected mutations increase in frequency in the population alongside with "swept" neutral or nearly neutral mutations (Smith and Haigh 1974). Additionally, deleterious mutations will be dragged along with the advantageous ones, if beneficial fitness effects outweigh the harmful consequences. Genetic hitchhiking also results in a reduced level of polymorphism at neutral sites adjacent to variants targeted by selection. Background selection and the ratchet are only expected to be effective when there are many sites under selection. In contrast, genetic hitchhiking can contribute to decay when there are fewer sites under selection, provided there is strong and frequent positive selection (Bachtrog 2008). The relative importance of these processes and the rate of decay is therefore expected to depend on the genomic architecture and the age of the non-recombining region. For instance, genetic hitchhiking should be more important for degeneration of older supergenes that already experienced gene loss, whereas Muller's ratchet and background selection have a greater impact on younger supergenes with more intact genes (Bachtrog 2008). Furthermore, for all of these processes, the extent of recombination or gene conversion between alternative haplotypes of the supergene is expected to have an impact on evolutionary trajectories, reducing divergence and decay.

While theoretical models on background selection can give valuable insights into expected patterns of degeneration in non-recombining regions, these models can break down when there are many sites under selection (McVean and Charlesworth 2000). Specifically, levels of polymorphism in non-recombining regions can then be higher than predicted under background selection (Kaiser and Charlesworth 2009). In this scenario, termed the interference selection limit, forward simulations or more complex models are necessary to develop reliable intuition about the effects of selection on linked variants (Good et al. 2014). This could often be the case in low-recombining regions, such as at supergenes. In Box 2 we show an example of how forward simulations of supergenes can be used to contrast expected patterns of accumulation of deleterious alleles at supergenes of different sizes and harboring two different types of structural variation (hemizygous region/insertion vs inversion).

Suppression of recombination can also directly alter the effective population size ($N_e$) of a supergene. A reduction in recombination generates a pseudo-population substructure causing supergene haplotypes to behave like separate populations that exchange migrants. Frequency differences between recombining and non-recombining chromosomes or supergene haplotypes can further magnify the reduction in $N_e$. For every three copies of the X chromosome in the population, there is only one Y chromosome, such that the impact of genetic drift over selection is enhanced at the Y chromosome relative to X chromosomes and autosomes (reviewed by Charlesworth and Charlesworth 2000). Likewise, at the *Primula S*-locus that governs heterostyly, the $N_e$ of the dominant *S*-haplotype which experiences restricted recombination is reduced (to ¼ of other autosomal loci under equal morph frequencies). It would thus be expected to be subject to evolutionary forces similar to a non-recombining Y or W region of similar size (and density of selected sites, as explained above). This reasoning can be extended to other supergenes that determine mating type, which should also undergo accumulation of deleterious mutations (reviewed by Uyenoyama 2005). In this case the extent of $N_e$ reduction for each haplotype will depend on its equilibrium frequency (reviewed by Uyenoyama 2005).

However, frequencies of different mating types are typically stable, and accumulation of deleterious mutations should not strongly affect the frequency of alternative haplotypes of mating system supergenes. In situations where the frequency of alternative arrangements can vary, the accumulation of deleterious mutations lowers the marginal fitness leading to a reduction in frequency. This reduction in frequency further strengthens drift and weakens selection leading to a subsequent increase in the rate of deleterious mutation accumulation (Berdan et al. 2021). This feedback loop is broken in supergenes whose frequency is tightly regulated by other factors, e.g. negative frequency-dependent selection in the case of mating system supergenes. Thus, we may expect a difference in degradation between supergenes where mutation accumulation impacts frequency and supergenes whose frequency is determined by other factors.

The dominance of deleterious mutations at the supergene will play a major role for the level of degradation. In nascent sex chromosomes in animals, masking of recessive deleterious mutations by permanent heterozygosity may reduce the capacity of purifying selection to remove them (reviewed in Vicoso 2019). Conversely, in supergenes that are hemizygous or undergo haploid expression, masking of recessive mutations will be less important. This consideration could affect plant sex chromosome evolution, given that many genes are expressed in haploid gametophytic cells in plants (Hough et al. 2017), supergenes in social insects where males are haploid (Stolle et al. 2019), and supergenes containing hemizygous regions, such as the *Primula S*-locus (Li et al. 2016).

Regulatory changes that affect genes in non-recombining regions could also play an important part in genetic degeneration. For instance, if silencing of genes happens at an early stage (e.g. Sun et al. 2018), then further degeneration may result from genetic drift rather than interference among selected sites. Selective down-regulation of Y-linked genes that harbor deleterious alleles could also trigger a haploidization feedback loop that leads to degeneration (Lenormand et al. 2020; reviewed by Charlesworth and Charlesworth 2020). This process can be accelerated by selective interference when there are many sites under selection. Finally, intragenomic conflict could be an important driver of the evolution of Y chromosomes (Cocquet et al. 2012; reviewed by Bachtrog 2020).

## Genomic architectures of supergenes: implications for evolutionary trajectories

A detailed characterization of supergene genomic architecture is critical for better understanding the origin of complex adaptations governed by supergenes, and how this architecture shapes subsequent evolutionary trajectories. As discussed above, the expected evolutionary trajectories of non-recombining haplotypes depend on the number of selected sites (Bachtrog 2008), as well as on the type of structural variation present. The genomic revolution has finally enabled the study of supergene architecture (Box 1) and recent studies have shown that the underlying genomic architecture of complex balanced polymorphisms can vary greatly, and so do evolutionary genetic patterns (Table 1). Here, we illustrate variation in supergene genomic architectures and evolution based on a set of genomic studies that used cutting-edge tools to comprehensively characterize supergenes governing ant colony social form, mimicry in butterflies, and heterostyly in flowering plants. We chose these particular supergenes as examples, because they provide examples of the diversity of supergene architectures, molecular evolutionary patterns and modes of origin. The showcased supergenes thus include a social supergene with similarities to sex chromosomes in *Solenopsis,* a multi-site regulatory supergene in *Papilio*, a supergene with structural variation introduced by introgression in *Heliconius*, and finally a hemizygous supergene that governs heterostyly in *Primula*. For each supergene example, we review what is known about its origin, genomic architecture and evolution.

**Chromosome-scale suppression of recombination, multiple sequential inversions, and degenerative expansion at social supergenes in *Solenopsis* ants**

Red fire ants, *Solenopsis invicta*, harbor a social supergene which exhibits evolutionary patterns strikingly similar to those of nascent sex chromosomes. This supergene governs colony social organization into either multiple-queen (polygyne) or single-queen (monogyne) colonies, both of which are present in *S. invicta* (Figure 1A). The social polymorphism is inherited as a single Mendelian factor, which governs a large set of disparate morphological and behavioral traits associated with social form (Linksvayer et al. 2013). The first gene to be associated with social form was the odorant binding protein gene *Gp-9* (Keller and Ross 1998). *Gp*-9 was later found to be located within a "social supergene", consisting of a large ~13 Mb genomic region harboring two haplotypes termed *Sb* and *SB*, which do not recombine freely with each other (Wang et al. 2013).

The genotype at the social supergene determines whether workers will accept multiple fertile queens in the colony or not (Keller and Ross 1998; Linksvayer et al. 2013; Wang et al. 2013). In monogyne colonies, both queens and workers are *SB/SB* and do not accept multiple fertile queens. In polygyne colonies, queens that harbor the *Sb* allele are accepted, but those that are homozygous for the *SB* allele are killed upon reproductive maturity (Keller and Ross 1998). While workers can be either homozygous *SB/SB* or heterozygous *SB/Sb*, reproductive *Sb/Sb* queens are not found in this species (Wang et al. 2013), and all queens in polygynous colonies are heterozygous *SB/Sb*. Because *Sb* and *SB* do not recombine freely (Wang et al. 2013), recombination can only occur between *SB* haplotypes, but not between *Sb* haplotypes.

The *Solenopsis* social supergene is expected to exhibit evolutionary genetic similarities to young sex chromosome systems. Because, like Y-chromosomes, *Sb* haplotypes experience reduced recombination and effective population size, they should degenerate (Wang et al. 2013; Pracana et al. 2017; Stolle et al. 2019). In contrast, the *SB* haplotype should only exhibit a moderate reduction in diversity relative to autosomal levels, with the extent of the reduction depending on the frequency of monogyne and polygyne colonies and gene flow between them (Pracana et al. 2017).

*The* Sb *haplotype resembles a young sex chromosome*

A set of recent studies have characterized the origin, genomic architecture and evolution of the social supergene in detail (Wang et al. 2013; Pracana et al. 2017; Huang et al. 2018; Stolle et al. 2019; Yan et al. 2020). Phylogenomic analyses of six socially polymorphic *Solenopsis* species indicate that the social supergene likely arose once, ~500 kya (Wang et al. 2013; Stolle et al. 2019; Yan et al. 2020), around the divergence time of these six species (Yan et al. 2020).

The supergene is characterized by a large ~10 Mb inversion that includes *Gp-9* and more than 400 other protein-coding genes, and there are also two smaller inversions (Yan et al. 2020) (Figure 2A). At least two of these inversions disrupt protein-coding genes and affect gene expression, suggesting that the inversions may have had direct and immediate functional consequences (Wang et al. 2013, Huang et al. 2018, Yan et al. 2020). Analyses of divergence among the *SB* and *Sb* haplotypes, which is related to the time since recombination stopped, suggest that the three inversions were likely incorporated sequentially, with the largest fixing first, and the last inversion connecting the supergene to the centromeric region (Yan et al. 2020). Even though synonymous divergence between the *SB* and *Sb* haplotypes was generally low (<1%) and variable, the suggested relative timing of these inversions was supported by concordant patterns of divergence among the *SB* and *Sb* haplotypes at the three inversions in six socially polymorphic *Solenopsis* species. A sequential increase in the extent of the region of restricted recombination is similar to the evolutionary strata of some sex chromosomes, and suggests that there could have been selection for further suppression of recombination during the evolution of the social supergene.

Regarding patterns of molecular evolution, the social supergene does exhibit some evolutionary genetic similarities to sex chromosomes. For instance, the *Sb* haplotype harbors significantly reduced polymorphism (Yan et al. 2020), with massive reductions of polymorphism seen in some populations (Pracana et al. 2017). The *Sb* haplotype further seems to be accumulating repetitive elements in a process that has been termed “degenerative expansion” (Stolle et al. 2019), and previously documented for sex chromosomes (e.g.

papaya: Wang et al. 2012; Na et al. 2014 and the neo-Y chromosome of *Drosophila miranda*: Mahajan et al. 2018). While there is some evidence for elevated $d_N/d_S$ ratios on the *Sb* haplotype relative to levels outside of the social supergene (Yan et al. 2020), tentatively suggesting a lower efficacy of purifying selection, the lack of major loss-of-function mutations or gene expression loss suggests that there has not yet been rampant genic degeneration or silencing of genes on the *Sb* haplotype (Wang et al. 2013; Stolle et al. 2019). However, careful analyses have now begun to detect evidence of reduced expression of *Sb*-linked alleles and corresponding dosage compensation at *SB* alleles (Martinez-Ruiz et al. 2020).

There are several factors that might have attenuated the degeneration of the *Sb* haplotype. First, efficient purifying selection in haploid males could slow down decay (Stolle et al. 2019) if recessive mutations are important for degeneration. Second, it is possible that the social supergene in *Solenopsis* has not yet had time to fix many deleterious mutations, due to its relatively recent origin. Third, there is evidence for low levels of recombination (crossovers or gene conversion) between *SB* and *Sb*, which could also slow down decay (Yan et al. 2020). Finally, reproductive *Sb/Sb* queens occur in *S. richteri.* Even rare occurrence of such queens in other species could slow down degeneration overall, because recombination could occur at *Sb* in such individuals. Thus, while further forward simulations may be needed to fully quantify the relative contribution of recombination and/or gene conversion, reproductive *Sb*/*Sb* queens and haploid selection to limited degeneration, overall the evolutionary patterns at the social supergene in red fire ants adhere to expectations under classic models of supergene evolution.

**A multi-site supergene governs female-limited Batesian mimicry in *Papilio polytes***

Batesian mimics are species with no natural defenses that are instead protected against predation due to their resemblance of defended, toxic “model” species. Many species of swallowtail butterflies in the *Papilio* genus are polymorphic for female-limited Batesian mimicry. Females of these species exhibit either a mimetic or a non-mimetic pattern, whereas males are non-mimetic (Figure 1B). The mimetic resemblance is based on a combination of wing patterns, wing and body color, and sometimes even the presence or absence of hindwing tails, yet early crossing experiments showed that color patterns in polymorphic female-limited Batesian mimics were inherited as a single Mendelian locus (Clarke and Sheppard 1960). Because developmentally disparate traits are involved, a supergene architecture consisting of several tightly linked genes was promptly hypothesized to underlie this phenomenon (Fisher 1930; Clarke and Sheppard 1960).

*The* P. polytes *mimicry supergene resembles a molecular switch*

Recent genomic studies have shown that a single, large gene rather than several tightly linked genes controls the female mimetic morphs in several *Papilio* species (Kunte et al. 2014; Nishikawa et al. 2015; Iijima et al. 2018, 2019; Palmer and Kronforst 2020). In *P. polytes,* two early studies mapped the genetic basis of female-limited mimicry to a ~130 kb autosomal region harboring the *doublesex* (*dsx*) gene (Kunte et al. 2014; Nishikawa et al. 2015) (Figure 2B). This region contains two highly differentiated haplotypes, termed *H* and *h*, which are associated with the mimetic and non-mimetic forms, respectively, and of which *H* is dominant and likely derived (Kunte et al. 2014; Nishikawa et al. 2015). Linkage disequilibrium is high within the *dsx* gene relative to adjacent genomic regions (Kunte et al. 2014), suggesting restricted recombination in *dsx*. The two haplotypes also differ with respect to an inversion, the breakpoints of which flank *dsx*, potentially contributing to recombination suppression (Kunte et al. 2014; Nishikawa et al. 2015).

Through a series of elegant knock-down experiments, Nishikawa et al. (2015) showed that the *H* variant of *dsx* (termed *dsx(H)*) acts as a molecular switch that turns on a pre-determined mimetic wing color pattern and suppresses the non-mimetic color pattern. Indeed, in *P. polytes*, expression of *dsx(H)* is specifically upregulated at early stages in wings of mimetic females, and the expression of *dsx(H)* is in turn associated with up- and downregulation of a suite of genes (Iijima et al. 2019). The upregulated genes include *Wnt1* and *Wnt6*, whereas *abdominal-A* is repressed (Iijima et al. 2019). Functional work supports a role for the first two genes in determining the red and white pigments that create the mimetic pattern, whereas *abdominal-A* likely inhibits the production of such coloration (Iijima et al. 2019). The size of the mimetic pattern elements in mimetic females further depends on their genotype at the supergene, suggesting that there is a dosage effect of *dsx(H),* and expression analyses suggest that this could be mediated by its effects on the gene network that it regulates (Iijima et al. 2019).

It thus seems clear that *dsx* is a major switch that underlies female-limited mimicry in *P. polytes*. As *dsx* encodes a highly conserved transcription factor with an important role in sex determination, its role in controlling mimicry was initially surprising. However, modularity of the Dsx protein and expression of different isoforms in different tissues, developmental stages and sexes could facilitate the evolution of these dual roles (Kunte et al. 2014; Nishikawa et al. 2015; Iijima et al. 2019). The exact mechanism by which *dsx(H)* controls mimetic color patterning is not known, but could involve protein-coding changes or a suite of *cis*-regulatory changes that increased expression of *dsx(H)* specifically in females (Iijima et al. 2019).

The key involvement of a regulatory gene in *Papilio* mimicry is in line with Nijhout's (1994) proposal that butterfly mimicry supergenes may often involve regulatory genes with quantitative effects on the expression of downstream structural genes. The genomic architecture of female-limited mimicry in *P. polytes* further illustrates what Booker et al. (2015) termed the "multi-site supergene model". Despite the resemblance to a molecular switch, this supergene may have evolved in multiple small steps similar to sex chromosomes, rather than by a single large-effect mutation. This would be in line with Clarke and Sheppard's (1960) suggestion that "it is probable that the switch mechanism itself evolved by a series of small steps" (Clarke and Sheppard 1960). In addition, morph-specific modifiers improving mimetic resemblance could also have been important (reviewed by Booker et al. 2015; Charlesworth 2016a).

While these studies have greatly contributed to an improved mechanistic understanding of how *dsx* governs mimicry, we know less about patterns of degeneration at this supergene. It is not clear that rapid decay would be expected, as the limited size of the supergene limits the efficacy of several processes contributing to decay, and the functional importance of *dsx* potentially poses strong constraints for accumulation of mutations. Further simulation-based work such as those shown in Box 2 would be valuable to explore expected patterns of mutation accumulation in this supergene.

*Distinct* dsx *alleles contribute to female-limited mimicry in different* Papilio *species*

While the first in-depth studies of the genetic basis of female-limited Batesian mimicry in *P. polytes* suggested that recombination suppression might be due to the presence of the inversion flanking *dsx*, subsequent studies suggest that this may not generally be the case in other *Papilio* species. For instance, in *Papilio memnon*, a close relative of *P. polytes*, polymorphic female-limited mimicry also maps to a region that contains *dsx* (Komata et al. 2016). This region also exhibits elevated nucleotide polymorphism and repeat content as well as strong linkage disequilibrium, suggestive of recombination suppression. Yet there is no evidence for an inversion in *P. memnon* (Iijima et al. 2018). These results suggest that recombination suppression could have predated the establishment of the inversion polymorphism in *P. polytes*.

Comparative genomic studies have further elucidated the role of *dsx* for female-limited mimicry across the *Papilio* genus. While *dsx* has been implicated in at least four cases of female-limited mimicry in *Papilio*, different species show independently derived mimicry alleles at *dsx*, and patterns of molecular differentiation ($F_{ST}$) between morphs and linkage disequilibrium in the region vary greatly (Komata et al. 2016; Palmer and Kronforst 2020). This complex pattern of variation at the mimicry supergene could be a result of repeated selection on variation at *dsx* and turnover of *dsx* alleles in different *Papilio* lineages (Palmer and Kronforst 2020). These surprising findings on the evolution of female-limited Batesian mimicry in swallowtail butterflies highlight the importance of studying supergene origins and evolution in a comparative framework.

**Introgression and chromosomal rearrangements underlie multiple mimetic morphs in *Heliconius numata***

Many *Heliconius* species exhibit Müllerian mimicry, a form of mimicry where well-defended species resemble each other's coloration patterns to achieve improved protection against predators. Studies on the genetic basis of Müllerian mimicry within this lineage show that the introgression of genetic structural variants has been associated with major adaptive novelties throughout its evolutionary history (Jay et al. 2018; Edelman et al. 2019). Some of the most in-depth insights on this phenomenon derive from studies on the supergene that governs alternative warning coloration patterns in *Heliconius numata*, which presents up to seven different sympatric morphs in populations distributed throughout the Amazonian basin and the Andean foothills (Joron et al. 2006).

*Introgression of an inverted haplotype contributed to the origin of the* H. numata *mimicry supergene*

Initial genetic work determined that the complex mimicry polymorphism in *H. numata* is governed by the *P* locus, a genomic region that harbors a set of genomic rearrangements that guarantee the joint inheritance of loci involved in color patterning due to the presence of a 400-Kb inversion ($P_1$) containing 21 genes (Joron et al. 2006, 2011; Jay et al. 2021). The $P_1$ inversion includes *cortex*, a gene involved in pigmentation patterning in many butterfly lineages (Nadeau et al. 2016). Individuals homozygous for the ancestral recessive allele (*Hn0*) display the *silvana* morph, while two different dominance relationships between derived alleles determine the phenotype in the other genotypes. While all derived haplotypes show complete dominance over the ancestral haplotype, for any combination of derived haplotypes the phenotype of wing patterning depends of the hierarchy in color expression within alleles of the derived class (Le Poul et al. 2014). Experiments suggest that polymorphism in *H. numata* is maintained by antagonistic frequency-dependent selection (Chouteau et al. 2017). Specifically, positive frequency-dependent selection imposed by predators increases the frequency of more common and better protected morphs, whereas disassortative mating counteracts the fixation of the underlying alleles (Chouteau et al. 2017).

Recent genomic studies have elucidated the complex origin, genomic architecture and selective forces behind the evolution of the *H. numata P* locus. Whole-genome analyses of species in the Silvaniform clade, to which this species belongs, showed that the inversion that contributed to the origin of the $P_1$ allele is fixed in the distantly related and sympatric *H. pardalinus* but is absent in other closely related species. There is also an excess of shared derived mutations between *H. numata* and *H. pardalinus* within this region, and these two species diverged long before the $P_1$ alleles they carry. Together, these findings indicate that the $P_1$ allele was introgressed into the *H. numata* population from *H. pardalinus* (Jay et al. 2018). This result suggests that introgression of the inverted haplotype contributed to the initial establishment of the supergene, without the evolution of close linkage in response to interactions among loci. Once established in *H. numata*, the $P_1$-carrying allele not only spread but also diversified through the occurrence of two sequential chromosomal rearrangements known as $P_2$ (200 Kb, 15 genes) and $P_3$ (1,150 Kb, 71 genes) (Jay et al. 2018, 2019) (Figure 2C).

*Inverted regions of the P locus supergene exhibit signatures of degeneration*

Several lines of evidence suggest that purifying selection may be weaker in the inverted arrangement compared to the standard arrangement of the *P* locus. For instance, the inverted regions have experienced a recent accumulation of transposable elements (TEs), which contributed to an ~9% size increase compared to the non-inverted regions on the ancestral *Hn0* arrangement (Jay et al. 2021). Furthermore, Jay et al. (2021) found that the inversions $P_1$, $P_2$ and $P_3$ show an elevated proportion of nonsynonymous relative to synonymous polymorphism ($p_N/p_S$) in comparison to genome-wide estimates and estimates for non-inverted homologous regions. Interestingly, experiments have shown strongly reduced larval survivorship of individuals homozygous for the same derived inverted arrangement at the supergene (Jay et al. 2021). Thus, the accumulation of mutations in *H. numata* may also contribute to maintenance of the supergene polymorphism, as heterozygotes for alternative recessive deleterious mutations are fitter than homozygotes. This pattern could constitute an example of associative overdominance resulting from linkage disequilibrium between derived structural variants and recessive deleterious alleles. This would be in line with Ohta's prediction that associative overdominance would be especially likely to be responsible for heterozygote superiority in the case of chromosomal inversions (Ohta 1971). The role of associative overdominance in the maintenance of supergene polymorphisms remains a major outstanding question.

It is still unclear whether the $P_2$ and $P_3$ rearrangements are a cause or a consequence of suppressed recombination. If selection against additional structural variants becomes less efficient in non-recombining haplotypes, such rearrangements could accumulate neutrally. As for the origin of $P_1$, evidence indicates that the inversion occurred in *H. pardalinus* and was subsequently introgressed into *H. numata* (Jay et al. 2018), suggesting that this inversion is not a consequence of suppressed recombination. The *H. numata P* supergene therefore constitutes an interesting example where the introgression of an inverted haplotype contributed to the evolution of a supergene.

### A supergene with a large hemizygous region governs heterostyly in *Primula*

Heterostyly is a floral adaptation to promote outcrossing that has fascinated many generations of evolutionary biologists, including Darwin (Darwin 1862, 1877). It is found in 28 flowering plant families (Barrett 2002a), having evolved independently at least 23 times (Lloyd and Webb 1992), and constitutes a compelling example of convergent evolution in plants (Ganders 1979). Herkogamy, the spatial separation of male and female reproductive organs (anthers and stigma), is common in flowering plants. Heterostylous species are special in that individuals present one of two or three types of flowers that differ reciprocally in the positioning of male and female reproductive organs within the flower, they exhibit reciprocal herkogamy. Among heterostylous species, those having two different types of flowers are known as distylous: the so-called L-morph (often referred to as "pin") has long styles and anthers positioned at the base of the floral tube, while S-morph (or "thrum") individuals present flowers with short styles and anthers at a high position in the flower (Figure 1C). Additional differences between morphs, e.g. in the size of pollen grains and the surface structure of the stigma, can also occur (Dulberger 1992) and may have functional significance (Costa et al. 2017). This floral polymorphism has frequently evolved in association with a heteromorphic self-incompatibility (SI) system that prevents successful self- and intra-morph pollination, such that L-morph individuals can only fertilize S-morph individuals and *vice versa.* In heterostylous plants, reciprocal herkogamy is thought to be beneficial because it promotes efficient pollen transfer and reduces sexual interference (Barrett 2002b), whereas SI allows for inbreeding avoidance (reviewed in Barrett 2019).

Bateson and Gregory (1905) showed early on that distyly is inherited as a single diallelic Mendelian locus, with the short style allele being dominant over the long style allele. The classic model for the distyly supergene, termed the *S*-locus, was conceived by Alfred Ernst (1936), who proposed that distyly was governed by a set of tightly linked loci with separable effects on different aspects of floral morph type and incompatibility. He suggested a model for the *S*-locus that included at least three separate loci governing style length and female incompatibility (*G*), anther position (*A*) and pollen size and male incompatibility (*P*). Under this model, S-morph plants were thus heterozygotes (*S/s,* or *GPA/gpa*) and L-morph plants were homozygous recessives at the *S*-locus (*s/s*, or *gpa/gpa*). Ernst considered floral morphs with intermediate character combinations to be the result of mutation, but rare recombination was later accepted as a more likely cause of breakdown of distyly to homostyly (Dowrick 1956; Lewis and Jones 1992). A third explanation posits that modifiers unlinked to the supergene might gradually alter style and stamen position (Mather and Winton 1941; reviewed by Ganders 1979). Classic genetic work in *Primula* thus suggested that heterostyly was governed by a diallelic supergene consisting of closely linked loci with separable effects on different aspects of morph differences and incompatibility, the traits of the S-morph being governed by dominant alleles of the underlying genes (reviewed by Kappel et al. 2017).

*The distyly supergene in* Primula *harbors a large hemizygous region formed by gene duplication and translocation*

Despite the long-standing interest in the genetic basis of distyly, until recently very little was known about the genomic architecture of the *S*-locus. This is now rapidly changing, thanks to genomic and functional work on the *Primula S*-locus (Huu et al. 2016; Li et al. 2016; Burrows and McCubbin 2017; Cocker et al. 2018). The recent sequencing of the *S*-locus in *Primula* showed that the dominant *S*-haplotype harbors a 278-kb insertion relative to the recessive *s*-haplotype (Li et al. 2016; Cocker et al. 2018) (Figure 2D). The insertion harbors five genes, including *CYP734A50,* which likely governs style length, based on functional studies (Huu et al. 2016) as well as sequence and expression analyses of naturally occurring floral variants (homostyles; Huu et al. 2016; Li et al. 2016). Another gene within the 278-kb insertion, *GLO*$^{T}$ (also called *GLO2*) is involved in determining anther position (Li et al. 2016; Huu et al. 2020). Thus, the 278-kb insertion in the dominant *S*-haplotype harbors at least two genes important for the determination of morph differences, and characterization of homostylous mutants has demonstrated that breakdown of distyly is not due to recombination but due to mutations, as originally proposed by Ernst (Huu et al. 2016; Li et al. 2016). The *S*-locus presence-absence polymorphism is shared among distylous species that diverged more than 20 Mya (Huu et al. 2016) and analyses of gene duplication timing suggest that the *S*-locus may be as old as 50 My, predating the origin of heterostyly in *Primula* (Li et al. 2016). *S*-locus polymorphism has therefore been maintained over extended evolutionary timescales, as one would expect for a locus under strong long-term balancing selection.

The presence of a large polymorphic insertion governing distyly potentially offers an explanation both for the rarity of recombination at the *S*-locus, and the dominance of *S*-alleles, whose recessive counterparts are simply missing (Huu et al. 2016; Li et al. 2016). However, there is some uncertainty regarding the degree of recombination suppression around the *S*-locus (Kappel et al. 2017), which is located next to a centromere (Li et al. 2015) and thus might be expected to be located in a genomic region with generally low recombination rates. Thus, it is not entirely clear whether the insertion caused recombination suppression at the supergene, or whether recombination was ancestrally low in this genomic region. Two of the constituent genes in the hemizygous region, *CYP734A50* and *GLO*$^{T}$, have paralogs outside of the *S*-locus (Huu et al. 2016; Li et al. 2016; Burrows and McCubbin 2017). This indicates that the *Primula S*-locus could be an example of a supergene that originated by gene duplication and translocation driven by selection, possibly involving interactions. A recent study showed that the paralogs of *CYP734A50* and *GLO*$^{T}$ are unlinked or at least very distant on the same chromosome (Huu et al. 2020), supporting stepwise assembly of the *S*-locus. Molecular evolutionary analyses of these two genes and their paralogs did not permit firm conclusions on which of these genes, $GLO^T$ or *CYP734A50,* was duplicated first (Huu et al. 2020). Still, stepwise duplications seem more likely than an origin through one large segmental duplication followed by gene loss and neofunctionalization, as previously suggested (Kappel et al. 2017).

*Limited molecular genetic evidence for degeneration at the* Primula *S-locus*

The genomic architecture of the distyly *S*-locus is expected to affect patterns of molecular evolution. As only crosses between S- and L-morph plants result in offspring, no *S/S* genotypes are generated, and the dominant *S*-haplotype is expected to experience a reduction in its $N_e$ to one quarter of that at autosomal loci. Lack of recombination in combination with reduced $N_e$ is expected to lead the dominant *S*-haplotype to accumulate deleterious alleles and repeats at a higher rate than collinear autosomal regions (Box 2). However, due to hemizygosity, selection on recessive alleles (either beneficial or deleterious) is expected to be more efficient in the *S*-haplotype, countering the effects outlined above. This is expected to slow down degeneration of the dominant *S*-haplotype, relative to expectations for supergenes that harbor inversions (Box 2).

Genetic studies have documented segregation patterns consistent with the existence of thrum-linked recessive lethal alleles (Kurian and Richards, 1997), implying that the dominant *S*-haplotype may be accumulating recessive deleterious alleles. As most molecular genetic studies have focused on gene function and genomic architecture, it is still unclear whether the dominant *S*-haplotype shows evolutionary genetic signatures of decay. However, Huu et al. (2016) showed that *CYP734A50* is evolving significantly faster than its non-*S*-linked paralog, likely as a result of a lower efficacy of purifying selection. The *Primula vulgaris S*-locus further exhibits an excess of repeats and TE-derived sequences relative to the genome-wide average values for assembled contigs, and relative to 171 kb of immediately flanking assembled regions (Cocker et al. 2018). Further assessment of variation in repeat content across *Primula* chromosomes using more contiguous genome assemblies would be useful to assess how unusual the repeat content of the *S*-locus region is, and further molecular evolutionary analyses to test for degeneration are also warranted.

In conclusion, recent genomic studies have shown that the *Primula S*-locus genomic architecture differs significantly from expectations under the classic model of a diallelic supergene, and that a hemizygous region found only on the dominant *S*-haplotype contains genes that are functionally important for distyly. The genomic architecture of the distyly supergene in *Primula* differs markedly from supergene architectures in *Solenopsis* and *Papilio* with respect to the type of structural variation it harbors (Figure 2), and more work will be required to understand whether the *S*-locus is degenerating.

Elucidation of distyly supergenes is now underway in other systems with similar inheritance of distyly, including *Linum, Fagopyrum,* and *Turnera* (Ushijima et al. 2012; Yasui et al. 2012; Shore et al. 2019, see also review by Kappel et al. 2017). Interestingly, in several of these systems hemizygous *S*-linked regions have been found, suggesting that hemizygosity could be a general feature of distyly *S*-loci (discussed in Kappel et al. 2017; Barrett 2019). However, in at least one system (*Turnera*) the dominant *S*-haplotype harbors several derived inversions in addition to an insertion (Shore et al. 2019), suggesting that it is premature to completely rule out a role for inversions in the evolution of heterostyly *S*-loci in plants. These findings on heterostyly supergenes thus highlight how new genomic studies can revolutionize our understanding of model systems in classical genetics.

## Conclusions and open questions

In this review, we have showcased recent genomic studies that have provided novel insights into the link between genomic architectures and the evolutionary fate of supergenes. So far, the clearest evidence for degeneration have been found at large supergenes harboring inversions, as expected from theory and simulations. Using simulations, we also showed that the genomic architecture, and in particular the type of structural variation present at a supergene, is crucial for the rate of degeneration.

An outstanding question concerns whether the processes that lead to accumulation of repeats and deleterious mutations could also contribute to maintenance of supergene polymorphisms. Both theoretical and empirical work indicate a potentially critical role for associative overdominance. Recent theoretical work has shown that a transition from background selection to associative overdominance can occur in regions of low recombination (Gilbert et al. 2020). Empirically, in the case of Müllerian mimicry of *H. numata,* it has been proposed that disassortative mating among different mimetic morphs evolved to avoid negative fitness consequences resulting from the expression of recessive deleterious load in homokaryotypes for derived inverted arrangements (Jay et al. 2021).

Recessive lethality of derived inverted supergene haplotypes has also been documented in the ruff supergene that governs mating morphs (Küpper et al. 2016). Could it therefore be that deleterious mutation accumulation contributes to balancing selection maintaining supergene architectures? The evolution of the genetic load at supergenes and the associated evolutionary consequences deserve to be investigated in more depth along with processes that involve adaptive variation.

Another outstanding question concerns the evolution of recombination suppression at supergenes other than sex chromosomes. How often does it occur, and what are the driving forces? It remains unclear what forces drive cessation of recombination around supergenes, and what determines whether the non-recombining region expands or not. Investigating the nature and impact of selective and neutral processes in shaping the evolution of supergene haplotypes is an important aim for future studies. Furthermore, determining the extent of the reduction in recombination and the context in which it occurs (i.e. only in heterokaryotypes vs. in all genotypes) will be key to linking this work with theoretical investigations of supergene degradation.

Theoretical and empirical work shows that the genomic architecture of a supergene is inextricably tied to its evolutionary fate. Here we have highlighted both empirical and theoretical techniques that will be crucial for further studies of this topic. Specifically, forward simulations (Box 2) and long read sequencing combined with new bioinformatic techniques (Box 1) provide avenues forward. Such work will allow for a more general understanding of the evolution of supergene architectures and the evolutionary causes and consequences of structural genomic variation.

## Acknowledgements

The authors thank the editor and four anonymous reviewers for comments that helped improve the manuscript, Alireza Foroozani and Yannick Wurm for comments on an early version of the manuscript, and Yannick Wurm, Krushnamegh Kunte, Haruhiko Fujiwara, and Sanford Porter for sharing photographs. This project has received funding from the European Research Council (ERC) under the European Union’s Horizon 2020 research and innovation programme (grant agreement No 757451) and from the Swedish Research Council (grant no. 2019-04452) to T.S. Emma Berdan was funded through a Carl Tryggers grant awarded to T.S. Computational work was enabled by resources provided by the Swedish National

Infrastructure for Computing (SNIC) at UPPMAX partially funded by the Swedish Research Council through grant agreement no. 2016-07213.

## Author Contributions

T.S. designed the research. T.S., J.G.V., P.W.H. and E.L.B. conducted literature review. E.L.B. conducted and analyzed the forward population genetic simulations. T.S. and J.G.V. wrote the first draft of the article. All authors edited and commented on the final manuscript.

## Literature Cited

Almeida P et al. 2020. Genome assembly of the basket willow, *Salix viminalis*, reveals earliest stages of sex chromosome expansion. BMC Biol. 18:78.

Arsenault SV et al. 2020. Simple inheritance, complex regulation: supergene-mediated fire ant queen polymorphism. Mol Ecol. 29:3622-3636.

Avril A, Purcell J, Béniguel S, Chapuisat M. 2020. Maternal effect killing by a supergene controlling ant social organization. Proc Nat Acad Sci USA. 117(29):17130–17134.

Bachtrog D. 2008. The temporal dynamics of processes underlying Y chromosome degeneration. Genetics. 179:1513-1525.

Bachtrog D. 2013. Y-chromosome evolution: emerging insights into processes of Y-chromosome degeneration. Nat Rev Genet. 14:113-124.

Bachtrog D. 2020. The Y chromosome as a battleground for intragenomic conflict. Trends Genet. 36:510-522.

Barrett SCH. 2002a. The evolution of plant sexual diversity. Nat Rev Genet. 3:274-284.

Barrett SCH. 2002b. Sexual interference of the floral kind. Heredity. 88:154-159.

Barrett SCH. 2019. ‘A most complex marriage arrangement’: recent advances on heterostyly and unresolved questions. New Phytol. 224:1051-1067.

Barton NH. 1995. Linkage and the limits to natural selection. Genetics. 140:821-841.

Bateson W, Gregory RP. 1905. On the inheritance of heterostylism in *Primula*. Proc R Soc Bio Sci. 76:581-586.

Berdan EL, Blanckaert A, Butlin RK, Bank C. 2021. Deleterious mutation accumulation and the long-term fate of chromosomal inversions. PloS Genet. In press.

Bergero R, Gardner J, Bader B, Yong L, Charlesworth D. 2019. Exaggerated heterochiasmy in a fish with sex-linked male coloration polymorphisms. Proc Natl Acad Sci USA. 116:6924-6931.

Booker T, Ness RW, Charlesworth D. 2015. Molecular evolution: breakthroughs and mysteries in Batesian mimicry. Curr Biol. 25:R506-508.

Branco S et al. 2017. Evolutionary strata on young mating-type chromosomes despite the lack of sexual antagonism. Proc Natl Acad Sci USA. 114:7067-7072.

Branco S et al. 2018. Multiple convergent supergene evolution events in mating-type chromosomes. Nat Commun. 9:2000.

Brelsford A et al. 2020. An ancient and eroded social supergene is widespread across *Formica* ants. Curr Biol. 30:304-311.

Burrows BA, McCubbin AG. 2017. Sequencing the genomic regions flanking *S*-linked *PvGLO* sequences confirms the presence of two *GLO* loci, one of which lies adjacent to the style-length determinant gene *CYP734A50*. Plant Reprod. 30:53-67.

Chakraborty M et al. 2018. Hidden genetic variation shapes the structure of functional elements in *Drosophila*. Nature Genetics. 50:20-25.

Charlesworth B, Charlesworth D. 2010. Elements of evolutionary genetics. Roberts and Company Publishers, Greenwood Village, CO.

Charlesworth B, Barton N. 2018. The spread of an inversion with migration and selection. Genetics. 208: 377-382

Charlesworth B, Charlesworth D. 1978. A model for the evolution of dioecy and gynodioecy. Am Nat. 112:975-997.

Charlesworth B, Charlesworth D. 2000. The degeneration of Y chromosomes. Philos Trans R Soc B. 355:1563-1572.

Charlesworth B, Charlesworth D. 2020. Evolution: a new idea about the degeneration of Y and W chromosomes. Curr Biol. 30:R871-R873.

Charlesworth B, Morgan MT, Charlesworth D. 1993. The effect of deleterious mutations on neutral molecular variation. Genetics. 134:1289-1303.
Charlesworth D, Charlesworth B. 1976. Theoretical genetics of Batesian mimicry II. Evolution of supergenes. J Theor Biol. 55:305-324.
Charlesworth D. 2016a. The status of supergenes in the 21st century: recombination suppression in Batesian mimicry and sex chromosomes and other complex adaptations. Evol Appl. 9:74-90.
Charlesworth D. 2016b. Plant sex chromosomes. Annu Rev Plant Biol. 67:397-420.
Charlesworth D. 2019. Young sex chromosomes in plants and animals. New Phytol. 224: 1095-1107.
Chouteau M, Llaurens V, Piron-Prunier F, Joron M. 2017. Polymorphism at a mimicry supergene maintained by opposing frequency-dependent selection pressures. Proc Natl Acad Sci USA. 114:8325-8329.
Christmas MJ et al. 2019. Chromosomal inversions associated with environmental adaptation in honeybees. Mol Ecol. 28:1358-1374.
Clarke CA, Sheppard PM. 1960. Super-genes and mimicry. Heredity. 14:175-185.
Cocquet J et al. 2012 A genetic basis for a postmeiotic X versus Y chromosome intragenomic conflict in the mouse. PLoS Genet. 8: e1002900.
Cocker JM et al. 2018. *Primula vulgaris* (primrose) genome assembly, annotation and gene expression, with comparative genomics on the heterostyly supergene. Sci Rep. 8:17942.
Costa J, Castro S, Loureiro J, Barrett SC. 2017. Experimental insights on the function of ancillary pollen and stigma polymorphisms in plants with heteromorphic incompatibility. Evolution. 71:121-134.
Crown KN, Miller DE, Sekelsky J, Hawley RS. 2018. Local inversion heterozygosity alters recombination throughout the genome. Curr Biol. 28:2984-2990.
Darlington CD, Mather K. 1949. The elements of genetics. George Allen & Unwin, London.
Darwin C. 1862. On the two forms, or dimorphic condition, in the species of *Primula*, and on their remarkable sexual relations. Bot J Linn Soc. 6:77-96.
Darwin C. 1877. The different forms of flowers on plants of the same species. John Murray, London.
Dobzhansky T, Epling C. 1948. The suppression of crossing over in inversion heterozygotes of *Drosophila pseudoobscura*. Proc Natl Acad Sci USA. 34:137-141.
Dobzhansky T, Sturtevant AH. 1938. Inversions in the chromosomes of *Drosophila pseudoobscura*. Genetics. 23:28-64.
Dowrick VPJ. 1956. Heterostyly and homostyly in *Primula obconica*. Heredity. 10:219-236.
Dulberger R. 1992. Floral polymorphisms and their functional significance in the heterostylous syndrome. In: Barrett SCH (ed) Evolution and function of heterostyly. Monographs on theoretical and applied genetics, vol. 15. Springer, Berlin.
Edelman NB et al. 2019. Genomic architecture and introgression shape a butterfly radiation. Science. 366:594-599.
Ernst A. 1936. Heterostylie-forschung. Versuche zur genetischen Analyse eines Organisations-und "Anpassungs" merkmales. Zeitschrift für Induktive Abstammungs- und Vererbungslehre. 71:156-230
Fisher RA. 1930. The genetical theory of natural selection. Clarendon Press, Oxford.
Fontaine MC et al. 2015. Mosquito genomics. Extensive introgression in a malaria vector species complex revealed by phylogenomics. Science. 347:1258524.
Fraser BA et al. 2020. Improved reference genome uncovers novel sex-linked regions in the guppy (*Poecilia reticulata*). Genome Biol Evol. 12:1789-1805.

Furman BLS et al. 2020. Sex chromosome evolution: so many exceptions to the rules. Genome Biol Evol. 12:750-763

Ganders FR. 1979. The biology of heterostyly. N Z J Botan. 17:607-635.

Gilbert KJ, Pouyet F, Excoffier L, Peischl S. 2020. Transition from background selection to associative overdominance promotes diversity in regions of low recombination. Curr. Biol. 30:101-107.

Good BH, Walczak AM, Neher RA, Desai MM. 2014. Genetic diversity in the interference selection limit. PLoS Genet. 10:e1004222.

Haller BC, Messer PW. 2019. SLiM 3: Forward genetic simulations beyond the Wright-Fisher model. Mol Biol Evol 36:632-637.

Horton BM, Michael CM, Prichard MR, Maney DL. 2020. Vasoactive intestinal peptide as a mediator of the effects of a supergene on social behavior. Proc R Soc Biol Sci. 287:20200196.

Hough J, Wang W, Barrett SCH, Wright SI. 2017. Hill-Robertson interference reduces genetic diversity on a young plant Y-chromosome. Genetics. 207:685-695.

Huang Y, Dang VD, Chang N, Wang J. 2018. Multiple large inversions and breakpoint rewiring of gene expression in the evolution of the fire ant social supergene. Proc R Soc Biol Sci. 285:20180221.

Huu CN et al. 2016. Presence versus absence of *CYP734A50* underlies the style-length dimorphism in primroses. eLife. 5:e17956.

Huu CN et al. 2020. Supergene evolution via stepwise duplications and neofunctionalization of a floral-organ identity gene. Proc Natl Acad Sci USA. 117: 23148-23157.

Iijima T et al. 2018. Parallel evolution of Batesian mimicry supergene in two *Papilio* butterflies, *P. polytes* and *P. memnon*. Sci Adv. 4:eaao5416.

Iijima T, Yoda S, Fujiwara H. 2019. The mimetic wing pattern of *Papilio polytes* butterflies is regulated by a *doublesex*-orchestrated gene network. Commun Biol. 2:257.

Jay P et al. 2018. Supergene evolution triggered by the introgression of a chromosomal inversion. Curr Biol. 28:1839-1845.e3.

Jay P et al. 2021. Mutation load at a mimicry supergene sheds new light on the evolution of inversion polymorphisms. Nature Genetics, In press. https://doi.org/10.1038/s41588-020-00771-1

Jordan CY, Charlesworth D. 2012. The potential for sexually antagonistic polymorphism in different genome regions. Evolution. 66:505-516.

Joron M et al. 2006. A conserved supergene locus controls colour pattern diversity in *Heliconius* butterflies. PLoS Biol. 4:e303.

Joron M et al. 2011. Chromosomal rearrangements maintain a polymorphic supergene controlling butterfly mimicry. Nature. 477:203-206.

Kaiser VB, Charlesworth B. 2009. The effects of deleterious mutations on evolution in non-recombining genomes. Trends Genet. 25:9-12.

Kappel C, Huu CN, Lenhard M. 2017. A short story gets longer: recent insights into the molecular basis of heterostyly. J Exp Bot. 68:5719-5730.

Keller L, Ross KG. 1998. Selfish genes: a green beard in the red fire ant. Nature. 394:573-575

Kim et al. 2017. A sex-linked supergene controls sperm morphology and swimming speed in a songbird. Nat Ecol Evol. 1:1168-1176.

Kirkpatrick M, Barton N. 2006. Chromosome inversions, local adaptation and speciation. Genetics. 173:419-434.

Komata S, Lin CP, Iijima T, Fujiwara H, Sota T. 2016. Identification of *doublesex* alleles associated with the female-limited Batesian mimicry polymorphism in *Papilio memno*n. Sci Rep. 6:34782.

Kuderna LFK et al. 2019. Selective single molecule sequencing and assembly of a human Y chromosome of African origin. Nat Commun. 10:4.
Kunte K et al. 2014. *doublesex* is a mimicry supergene. Nature. 507:229-232.
Kurian V, Richards AJ. 1997. A new recombinant in the heteromorphy 'S' supergene in Primula. Heredity 78:383-390.
Küpper C et al. 2016. A supergene determines highly divergent male reproductive morphs in the ruff. Nat Genet. 48:79-83.
Lahn BT, Page DC. 1999. Four evolutionary strata on the human X chromosome. Science 286: 964-967.
Lamichhaney S et al. 2016. Structural genomic changes underlie alternative reproductive strategies in the ruff (*Philomachus pugnax*). Nat Genet. 48:84-88.
Le Poul, Y et al. 2014. Evolution of dominance mechanisms at a butterfly mimicry supergene. Nat Commun. 5:5644.
Lenormand T, Fyon F, Sun E, Roze D. 2020. Sex chromosome degeneration by regulatory evolution. Curr Biol. 30:P3001-3006
Lewis D, Jones DA. 1992. The genetics of heterostyly. In: Barrett SCH (ed) Evolution and function of heterostyly. Monographs on theoretical and applied genetics, vol. 15. Springer, Berlin.
Li B et al. 2020. A large genomic insertion containing a duplicated *follistatin* gene is linked to the pea aphid male wing dimorphism. eLife. 9:e50608
Li J et al. 2016. Genetic architecture and evolution of the *S* locus supergene in *Primula vulgaris*. Nat Plants. 2:16188.
Li J et al. 2015. Integration of genetic and physical maps of the *Primula vulgaris* S locus and localization by chromosome *in situ* hybridization. New Phytol. 208:137-148.
Lieberman-Aiden E et al. 2009. Comprehensive mapping of long-range interactions reveals folding principles of the human genome. Science. 326:289-293.
Linksvayer TA, Busch JW, Smith CR. 2013. Social supergenes of superorganisms: do supergenes play important roles in social evolution? Bioessays. 35:683-689.
Lloyd DG, Webb CJ. 1992. The evolution of heterostyly. In: Barrett SCH (ed) Evolution and function of heterostyly. Monographs on theoretical and applied genetics, vol. 15. Springer, Berlin.
Mahajan S, Wei KH, Nalley MJ, Gibilisco L, Bachtrog D. 2018. De novo assembly of a young *Drosophila* Y chromosome using single-molecule sequencing and chromatin conformation capture. PLoS Biol. 16:e2006348.
Maney D, Merritt J, Prichard M, Horton B, Yi S. 2020. Inside the supergene of the bird with four sexes. Horm Behav. 126:104850.
Martinez-Ruiz C et al. 2020. Genomic architecture and evolutionary antagonism drive allelic expression bias in the social supergene of red fire ants. eLife 9:e55862.
Mather K, Winton DD. 1941. Adaptation and counter-adaptation of the breeding system in *Primula*. Ann Bot. 5:297-311.
McVean GA, Charlesworth B. 2000. The effects of Hill-Robertson interference between weakly selected mutations on patterns of molecular evolution and variation. Genetics. 155:929-944.
Merritt JR et al. 2020. A supergene-linked estrogen receptor drives alternative phenotypes in a polymorphic songbird. Proc Natl Acad Sci USA. 117:21673-21680.
Muller HJ. 1964. The relation of recombination to mutational advance. Mutat Res. 106:2-9.
Na JK, Wang J, Ming R. 2014. Accumulation of interspersed and sex-specific repeats in the non-recombining region of papaya sex chromosomes. BMC Genomics. 15:335.
Nadeau NJ et al. 2016. The gene *cortex* controls mimicry and crypsis in butterflies and moths. Nature. 534:106-110.

Navarro A, Betrán E, Barbadilla A, Ruiz A. 1997. Recombination and gene flux caused by gene conversion and crossing over in inversion heterokaryotypes. Genetics. 146:695-709.

Nijhout HF. 1994. Developmental perspectives on evolution of butterfly mimicry. BioScience. 44:148-157.

Nishikawa H et al. 2015. A genetic mechanism for female-limited Batesian mimicry in *Papilio* butterfly. Nat Genet. 47:405-409.

Nosil P et al. 2018. Natural selection and the predictability of evolution in *Timema* stick insects. Science. 359:765-770.

Ohta T. 1971. Associative overdominance caused by linked detrimental mutations. Genet Res. 18:277-86.

Palmer DH, Kronforst MR. 2020. A shared genetic basis of mimicry across swallowtail butterflies points to ancestral co-option of *doublesex*. Nat Commun. 11:6.

Palmer DH, Rogers TF, Dean R, Wright AE. 2019. How to identify sex chromosomes and their turnover. Mol Ecol. 28:4709-4724.

Pearse et al. 2019. Sex-dependent dominance maintains migration supergene in rainbow trout. Nat Ecol Evol. 3:1731-1742.

Pettersson ME et al. 2019. A chromosome-level assembly of the Atlantic herring genome-detection of a supergene and other signals of selection. Genome Res. 29:1919-1928.

Poorman PA, Moses MJ, Russell LB, Cacheiro NL. 1981. Synaptonemal complex analysis of mouse chromosomal rearrangements. I. Cytogenetic observations on a tandem duplication. Chromosoma. 81: 507-518.

Pracana R, Priyam A, Levantis I, Nichols RA, Wurm Y. 2017. The fire ant social chromosome supergene variant *Sb* shows low diversity but high divergence from *SB*. Mol Ecol. 26:2864-2879.

Purcell J, Brelsford A, Wurm Y, Perrin N, Chapuisat M. 2014. Convergent genetic architecture underlies social organization in ants. Curr Biol. 24:P2728-2732.

Rice WR. 1987. The accumulation of sexually antagonistic genes as a selective agent promoting the evolution of reduced recombination between primitive sex chromosomes. Evolution. 41:911-914.

Saenko SV et al. 2019. Unravelling the genes forming the wing pattern supergene in the polymorphic butterfly *Heliconius numata*. Evo Devo. 10:16.

Schwander T, Libbrecht R, Keller L. 2014. Supergenes and complex phenotypes. Curr Biol. 24:R288-294.

Shore JS et al. 2019. The long and short of the S-locus in *Turnera* (Passifloraceae). New Phytol. 224:1316-1329.

Smith JM, Haigh J. 1974. The hitch-hiking effect of a favourable gene. Genet Res. 23:23-35.

Stolle E et al. 2019. Degenerative expansion of a young supergene. Mol Biol Evol. 36:553-561.

Sturtevant AH, Beadle GW. 1936. The relations of inversions in the X chromosome of *Drosophila melanogaster* to crossing over and disjunction. Genetics 21: 554-604.

Sun D, Huh I, Zinzow-Kramer WM, Maney DL, Yi SV. 2018. Rapid regulatory evolution of a nonrecombining autosome linked to divergent behavioral phenotypes. Proc Natl Acad Sci USA. 115:2794-2799.

Sun Y, Svedberg J, Hiltunen M, Corcoran P, Johannesson H. 2017. Large-scale suppression of recombination predates genomic rearrangements in *Neurospora tetrasperma*. Nat Commun. 8:1140.

Thomas JW et al. 2008. The chromosomal polymorphism linked to variation in social behavior in the white-throated sparrow (*Zonotrichia albicollis*) is a complex rearrangement and suppressor of recombination. Genetics. 179:1455-1468.

Thompson MJ, Jiggins CD. 2014. Supergenes and their role in evolution. Heredity. 113:1-8.
Tomaszkiewicz M, Medvedev P, Makova KD. 2017. Y and W chromosome assemblies: approaches and discoveries. Trends Genet. 33:266-282.
Treangen TJ, Salzberg SL. 2011. Repetitive DNA and next-generation sequencing: computational challenges and solutions. Nat Rev Genet. 13:36-46.
Turner JRG. 1967. On supergenes. I. The evolution of supergenes. Am Nat 101: 195-221.
Tuttle EM et al. 2016. Divergence and functional degradation of a sex chromosome-like supergene. Curr Biol. 26:344-350.
Ushijima K et al. 2012. Isolation of the floral morph-related genes in heterostylous flax (*Linum grandiflorum*): the genetic polymorphism and the transcriptional and post-transcriptional regulations of the *S* locus. Plant J. 69:317-331.
Uyenoyama MK. 2005. Evolution under tight linkage to mating type. New Phytol. 165:63-70.
Vicoso B, Emerson JJ, Zektser Y, Mahajan S, Bachtrog D. 2013. Comparative sex chromosome genomics in snakes: differentiation, evolutionary strata, and lack of global dosage compensation. PLoS Biol. 11:e1001643.
Vicoso B. 2019. Molecular and evolutionary dynamics of animal sex-chromosome turnover. Nat Ecol Evol 3:1632-1641.
Villoutreix R et al. 2020. Large-scale mutation in the evolution of a gene complex for cryptic coloration. Science. 369:460-466.
Wang J et al. 2012. Sequencing papaya X and $Y^h$ chromosomes reveals molecular basis of incipient sex chromosome evolution. Proc Natl Acad Sci USA. 109:13710-13715.
Wang J et al. 2013. A Y-like social chromosome causes alternative colony organization in fire ants. Nature. 493:664-668.
Wright AE, Dean R, Zimmer F, Mank JE. 2016. How to make a sex chromosome. Nat Commun. 7:12087.
Wright S, Dobzhansky T. 1946. Genetics of natural populations. XII. Experimental reproduction of some of the changes caused by natural selection in certain populations of *Drosophila pseudoobscura*. Genetics 31: 125-156.
Yan Z et al. 2020. Evolution of a supergene that regulates a trans-species social polymorphism. Nat Ecol Evol. 4:240-249.
Yasui Y et al. 2012. S-LOCUS EARLY FLOWERING 3 is exclusively present in the genomes of short-styled buckwheat plants that exhibit heteromorphic self-incompatibility. PLoS ONE. 7:e31264.
Yeaman S. 2013. Genomic rearrangements and the evolution of clusters of locally adaptive loci. Proc Natl Acad Sci USA. 110:E1743-51.
Zheng G et al. 2016. Haplotyping germline and cancer genomes with high-throughput linked-read sequencing. Nat Biotechnol. 34:303-311.

Table 1. Supergenes identified using genomic methods, including information on the trait under selection, the type of selection maintaining polymorphism, the inferred age, size, gene content, identification strategy, and evolutionary genetic evidence for degeneration.

| **Trait (Locus)** | **Lineage** | **Origin** | **Structural variation** | **Selection** | **Age** | **Size (kb)** | **# genes in region** | **Candidate genes** | **ID strategy** | **Signs of degeneration** | **Refs** |
|---|---|---|---|---|---|---|---|---|---|---|---|
| Batesian mimicry wing color patterning (*H* locus) | *Papilio*<br>*1. polytes*<br>2. *memnon* | Multiple | 1. Inversion | Positive frequency-dependent selection | Unknown | 1. 130<br>2. 168 | 3 | *doublesex*, *Nach-like*, and *UXT* | 1. Genetic association mapping, morph-specific expression<br>2. Coverage, genetic differentiation | Accumulation of TEs and repetitive sequences | (Clarke and Sheppard 1960; Kunte et al. 2014; Nishikawa et al. 2015; Komata et al. 2016; Iijima et al. 2018) |
| Batesian mimicry wing color patterning (*P* locus) | *Heliconius numata* | Single | Inversion introduced by introgression | Antagonistic frequency-dependent selection | Inversion: 2.41 My, introgression: 2.30-2.24 My | P1=400<br>P2=200<br>P3=1,150 | P1=21<br>P2=15<br>P3=71 | *cortex* | Genetic linkage mapping, linkage disequilibrium (LD) analyses in natural populations and positional cloning | Accumulation of deleterious mutations and TEs, degenerative expansion | (Joron et al. 2006; Joron et al. 2011; Chouteau et al. 2017; Jay et al. 2018; Jay et al. 2019; Saenko et al. 2019) |
| Colony social organization (Social S-locus) | *Formica* | Single | Inversion | Maternal effect killer in *F. selysii* | 40-20 My | 11,000 | Varies | *Knockout*, serine-threonine kinase STK32B, MRPL34, *RPUSD4* and *G9A* | Genetic association mapping, genetic differentiation between haplotypes, morphotype – genotype association | No major evidence for degeneration, low differentiation between haplotypes except at clusters of *trans*-species SNPs | (Purcell et al. 2014; Avril et al. 2020; Brelsford et al. 2020) |
| Colony social organization (Social S-locus) | *Solenopsis* | Single | Two large inversions | *SB/SB* queens killed in polygyne colonies | 0.5 My | 13,000 | 616 | *Gp*-9 | Genetic association mapping, differential expression analyses | High frequency of deleterious mutations, repetitive elements, degenerative expansion | (Wang et al. 2013; Stolle et al. 2019; Yan et al. 2020; Arsenault et al. 2020) |
| Cryptic coloration morphs (*Mel-Stripe* locus; m, U, and S variants) | *Timema* | Single | Inversion | Balancing selection, spatial heterogeneity | Between m and U :13.5-8.0 My; between U and S: 2.7-1.8 My | 13,000 | 83 | A 1000 Kb deletion in this region controls coloration across several species | Genome-wide association study (GWAS), differences in read depth coverage | Not studied | (Nosil et al. 2018; Villoutreix et al. 2020) |
| Heterostyly (*S*-locus) | *Primula* | Single | Insertion | Disassortative mating, long-term balancing selection | 50 My | 278 | 5 | *CYP734A50* (style length), *GLO*$^{T}$*/GLO2* (anther position), *KFB*, *PUM* and *CCM* | Differentially expressed genes in specific floral organs of the two floral morphs, identification of *S*-linked loci, genetic and physical maps, comparison of *S* haplotype sequences, functional studies of both *CYP734A50* and *GLO*$^{T}$*/GLO2* | TEs accumulation: 64% in region vs. 37% genome wide | ( Huu et al. 2016; Li et al. 2016; Burrows and McCubbin 2017; Cocker et al. 2018; Huu et al. 2020) |
| Heterostyly (*S*-locus) | *Turnera* | Single | Three hemizygous genes + two inversions | Disassortative mating, long-term balancing selection | Unknown | 241 | 21 + 3 | *TsBAHD* (pistils), *TsSPH1 and TsYUC6* (stamens) | Deletion mapping to sequence BAC clones and genome scaffolds to construct haplotypes, organ-specific gene expression | Higher TE content in dominant *S*-haplotype than in recessive *s*-haplotype, two inversions (unclear if cause or consequence of suppressed recombination) | (Shore et al. 2019) |
| Male mating morphs | *Philomachus pugnax* | Single | Inversion | Balancing selection | 3.8 My | 4,400 | 25 | *HSD17B2*, *SDR42E1*, *ZDHHC7* and *CYB5B* (all | Genetic linkage mapping, GWAS, genetic sequence divergence analyses | Not studied | (Küpper et al. 2016; Lamichhan |
| | | | | | | | | involved in steroid metabolism) | | | ey et al. 2016) |
| Male-restricted dimorphism (*api* locus) | *Acyrthosiphon pisum* species complex | Single | Insertion | Balancing selection | 10 My | 120 | 12 | *follistatin* | QTL analysis, coverage differences and genetic sequence differentiation between morphs | Not studied | (Li et al. 2020) |
| Mating morphs (ZAL2/ZAL2m) | *Zonotrichia albicollis* | Single | Two inversions | Disassortative mating | 2.5-1.9 My | 100,000 | 1,137 | *ESR1* (aggressiveness) and *VIP* (aggressiveness, parental behavior), *FIG4* and *LYST* (pigmentation) | Comparative chromosome painting, cytogenetic mapping, genetic sequence diversity and divergence analyses, LD analyses in natural populations | Reduced genetic diversity, excess of nonsynonymous mutations, no substantial degeneration but gene expression changes | (Thomas et al. 2008; Tuttle et al. 2016; Sun et al. 2018; Maney et al. 2020; Horton et al. 2020; Merritt et al. 2020) |
| Mating type (MAT loci) | *Microbotryum* | Multiple | Chromosomal rearrangements, fusion of the MAT chromosomes | Balancing selection | Five independent times in the last 2.3-0.2My | 1,000-10,000 | 120-547 | Homeodomain transcription factor genes *PD* and *HD* loci, responsible for pre- and post-fertilization compatibility | Co-segregation of MAT type, chromosome dimorphism and markers; finding of contigs carrying *PD* and *HD*, comparative genomics (homology and synteny) | Gene losses, TEs accumulation | (Branco et al. 2017; Branco et al. 2018) |
| Rainbow trout migration | *Oncorhynchus mykiss* | Single | Two inversions | Sexually antagonistic balancing selection | 1.5 My | 5,500 | 1,091 | *DMRTA2, AMH, NR5A2, RORC1, RXRA, LEPR, CENPR, CLOCK,PDCL*, *PPEF2*, *RX3* and *MAPK10* (*JNK3*) | Genetic linkage mapping, genetic sequence diversity and divergence analyses | Not studied | (Pearse et al. 2019) |
| Sperm morphology | *Taeniopygia guttata* | Single | Z-linked inversion | Heterozygote advantage | Unknown | ~63,000 | 648 | *GADD45G*, *LRRC2*, *C9orf3*, *FBXL17*, *DMRT2*, *LINGO2*, *ZNF462, RAD23B* | GWAS, trait artificial selection, population genetic differentiation, expression quantitative trait locus (eQTL) analysis | Not studied | (Kim et al. 2017) |

**Box 1. Challenges and new methods to sequence supergenes**

Suppressed recombination is important for the formation of supergenes, but the genetic consequences of long-term suppressed recombination can also be an obstacle to the study of supergenes. In particular, the accumulation of repeats (see Degeneration of supergene haplotypes – theory and expectations) can complicate assembly of low-recombination regions (Tomaszkiewicz et al. 2017). Short-read whole-genome sequencing data can be especially problematic under these circumstances because repeats create ambiguities which result in assembly and read mapping errors (Treangen and Salzberg 2011). Although several bioinformatic methods have been developed to deal with these limitations (Treangen and Salzberg 2011), long-read sequencing technologies have drastically improved our ability to assemble highly repetitive regions.

Single-molecule and long-read sequencing data have been widely used for the assembly of complete and contiguous sex chromosome sequences (e.g. Kuderna et al. 2019; Almeida et al. 2020). High-quality genome assemblies enable successful identification of structural variants that remain hidden in more fragmented assemblies (Chakraborty et al. 2018). Indeed, highly contiguous genomes facilitated the characterization of large-scale inversions involved in ecological adaptation in honeybees (Christmas et al. 2019) and Atlantic herring (Pettersson et al. 2019). Long read-based assemblies can be more accurately phased, especially in combination with other high-throughput methods such as linked reads (Zheng et al. 2016) and chromosomal conformation and capture data (Hi-C) (Lieberman-Aiden et al. 2009). The production of phased haplotypes is of special interest for the study of supergenes.

Despite the difficulty of assembling regions under suppressed recombination, the genomic signatures of divergence and degeneration are regularly used for the identification of both supergenes and sex chromosomes. Palmer et al. (2019) recently reviewed bioinformatic strategies for genomic identification of sex chromosomes, and the same approaches can be successfully applied to study supergenes. The extent of divergence in particular determines the suitability of different strategies. For instance, nucleotide divergence and population differentiation can be useful to identify homomorphic supergenes (Tuttle et al. 2016; Sun et al. 2018), whereas repeat content analyses can facilitate detection of heteromorphic supergenes (Stolle et al. 2019), and read depth analyses have successfully identified supergenes with hemizygous regions (Li et al. 2016; Cocker et al. 2018; Li et al. 2020) (See Figure Box 1). Table 1 summarizes analyses used for the identification and study of several supergenes.

**Box 2. Forward simulations elucidate the impact of supergene architecture on patterns of molecular evolution.**

To illustrate how genomic architecture affects patterns of molecular evolution in supergenes we used SLiM v3.3.2 (Haller and Messer 2019), a forward simulation program, to model a hypothetical *S*-locus system harboring either an inversion or an indel (a region hemizygous in thrums). Both types of structural variation have been documented at *S*-loci (e.g. Li et al. 2016; Shore et al. 2019).

We modeled a population of 50,000 diploid individuals using parameter estimates from *Arabidopsis* to calibrate our model (see Supplementary Text for additional details on mutation and recombination rates and the distribution of selection coefficients). To recapitulate disassortative mating patterns and *S*-genotype distributions in distylous species, we only allowed matings between heterozygotes and homozygotes at the *S*-locus and considered *S/S* homozygotes lethal. We modeled the *S*-allele as an insertion (hemizygous case) or an inversion. For the hemizygous case, we examined how size affected fixation of mutations by varying the size of the region (0.625%, 1.25%, or 2.5% of the genome, corresponding to 31.25 kb, 62.5 kb, or 125 kb). The proportion of sites under selection was kept constant at 20% both at the *S*-locus and in the remaining genome simulated (total length 5 Mb). The vast majority of mutations (99.9%) were deleterious and recessive with magnitudes of fitness effects ($|s|$) drawn from a Gamma distribution $\Gamma$ ($\alpha$=0.5, mean=0.0025). Beneficial mutations (0.1% of mutations) were co-dominant and drawn from an exponential distribution with a mean of 0.001.

**Accumulation of mutations under selection**

An equal-sized 62.5 kb inversion accumulated >6x more deleterious mutations than a hemizygous region (Figure Box 2, panel A, B. Furthermore, these mutations had lower (i.e. more negative) selection coefficients (Figure Box 2, panels C-F). This is likely due to the fact that mutations in the *S*-haplotype insertion, but not in the *S*-haplotype with an inversion affect fitness in the heterozygous state. Larger insertions accumulated slightly more mutations (Figure Box 2, panel A) that were slightly more deleterious, but this effect was very subtle. There was no effect of either size or supergene type on accumulation of beneficial mutations under our simulation parameters (results not shown).

In our simulations, the supergene haplotype with suppressed recombination was held at a constant intermediate frequency (0.25) and always accumulated deleterious mutations. This accumulation varied little across runs as well as over the majority of the investigated parameter space. The largest factor affecting patterns of molecular evolution was whether the supergene harbored a hemizygous region or an inversion. Thus, supergenes where all deleterious mutations are exposed to selection in the heterozygous state may have a different evolutionary trajectory compared to supergenes where deleterious mutations are recessive or only partially dominant.

## Figure Legends

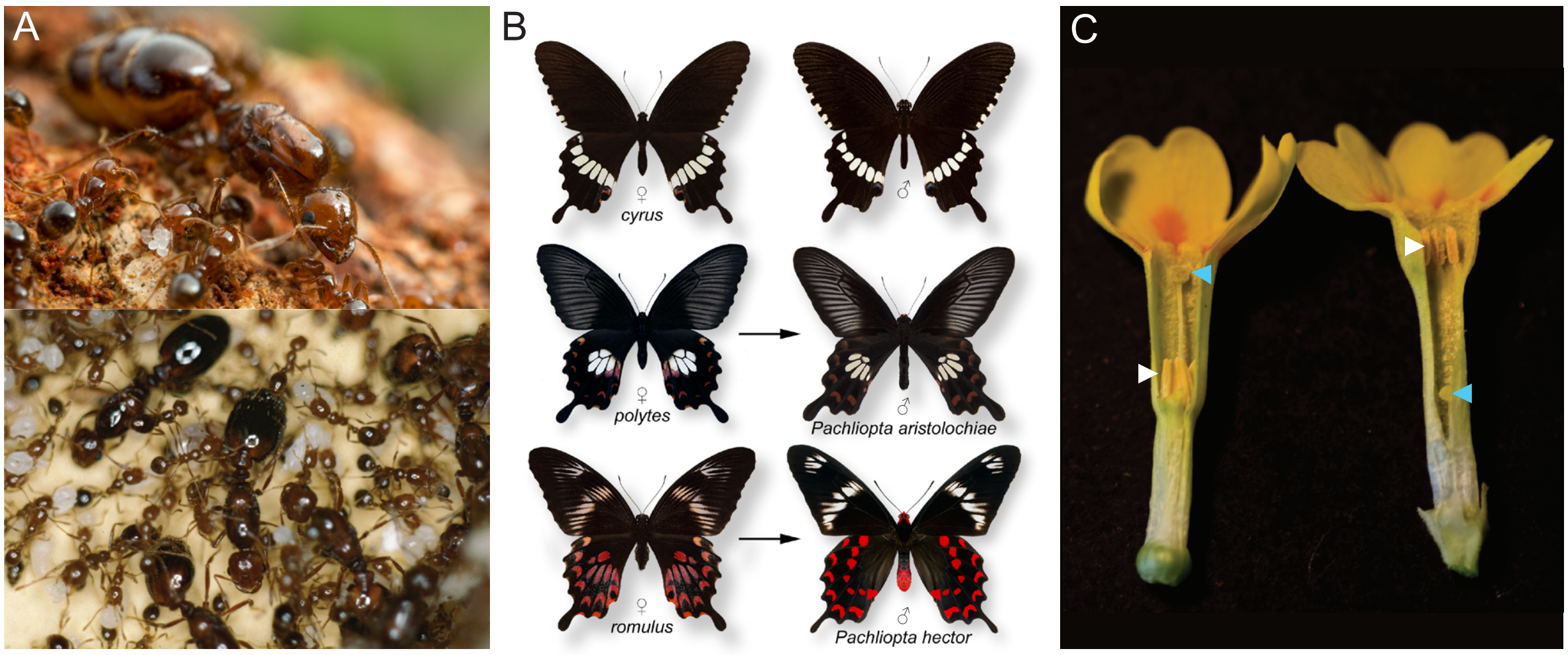

Figure 1. Balanced polymorphisms governed by supergenes. A. Colony social form (monogynous vs. polygynous) in *Solenopsis invicta*. Monogynous colonies have a single queen (top image, courtesy of Alex Wild) whereas polygynous colonies have multiple queens (bottom image, courtesy of SD Porter, USDA-ARS). B. Polymorphic female-limited Batesian mimicry in *Papilio polytes*. The non-mimetic female form *cyrus* and male *P. polytes* (top), the mimetic female form *polytes* and its model *Pachliopta aristolochiae* (male form), and the mimetic female form *romulus* and its model *Pachliopta hector* (male form). Image courtesy of Krushnamegh Kunte. C. Heterostyly in *Primula veris*. The pin morph (left) has the stigma in a high position (blue arrow) and anthers in a low position (white arrow) in the floral tube, whereas the thrum morph (right) has anthers in a high position (white arrow) and the stigma in a low position (blue arrow). Image courtesy of Tanja Slotte.

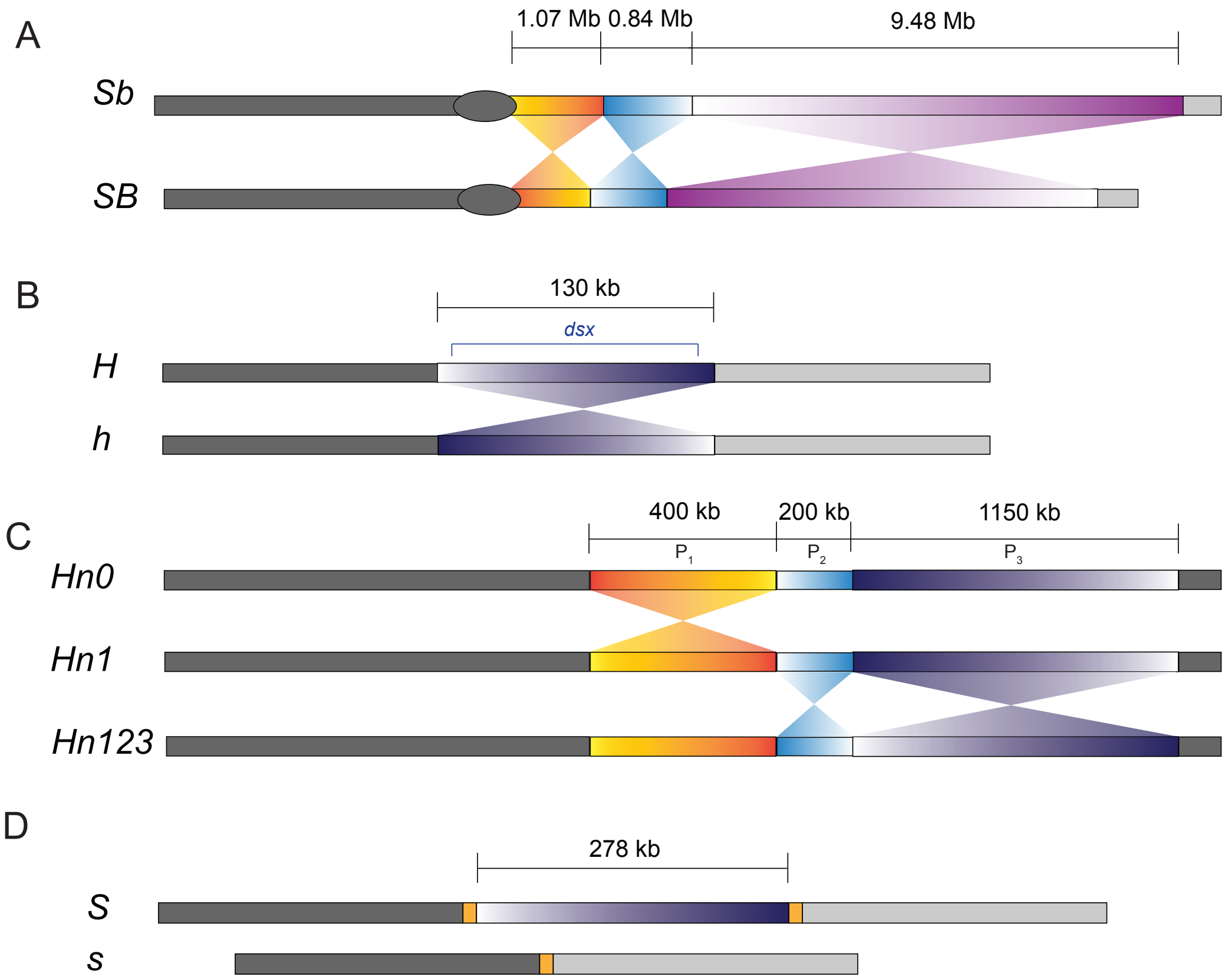

Figure 2. Differences in size and structure of four classic supergenes. A. The social supergene of *Solenopsis invicta*. The *Sb* haplotype harbors at least three large inversions relative to the *SB* haplotype and is longer than the *SB* haplotype due to repeat expansion (length of *SB* and *Sb* haplotypes not drawn to scale here). B. The female-limited mimicry supergene of *Papilio polytes.* The mimetic haplotype (*H*) harbors an ~130 kb inversion that flanks the gene *dsx* relative to the nom-mimetic haplotype (*h*). C. The *Heliconius numata* P supergene. The haplotype with the ancestral arrangement (*Hn0*) differs from the derived and more dominant haplotypes (*Hn1* and *Hn123*) with respect to a 400 kb inversion ($P_1$) introduced by introgression from *Heliconius pardalinus*. Haplotype *Hn123* harbors two additional derived inversions ($P_2$ and $P_3$) relative to both *Hn0* and *Hn1*. D. The *Primula vulgaris S*-locus contains a 278 kb hemizygous region present only on the dominant *S*-haplotype and not on the recessive *s*-haplotype.

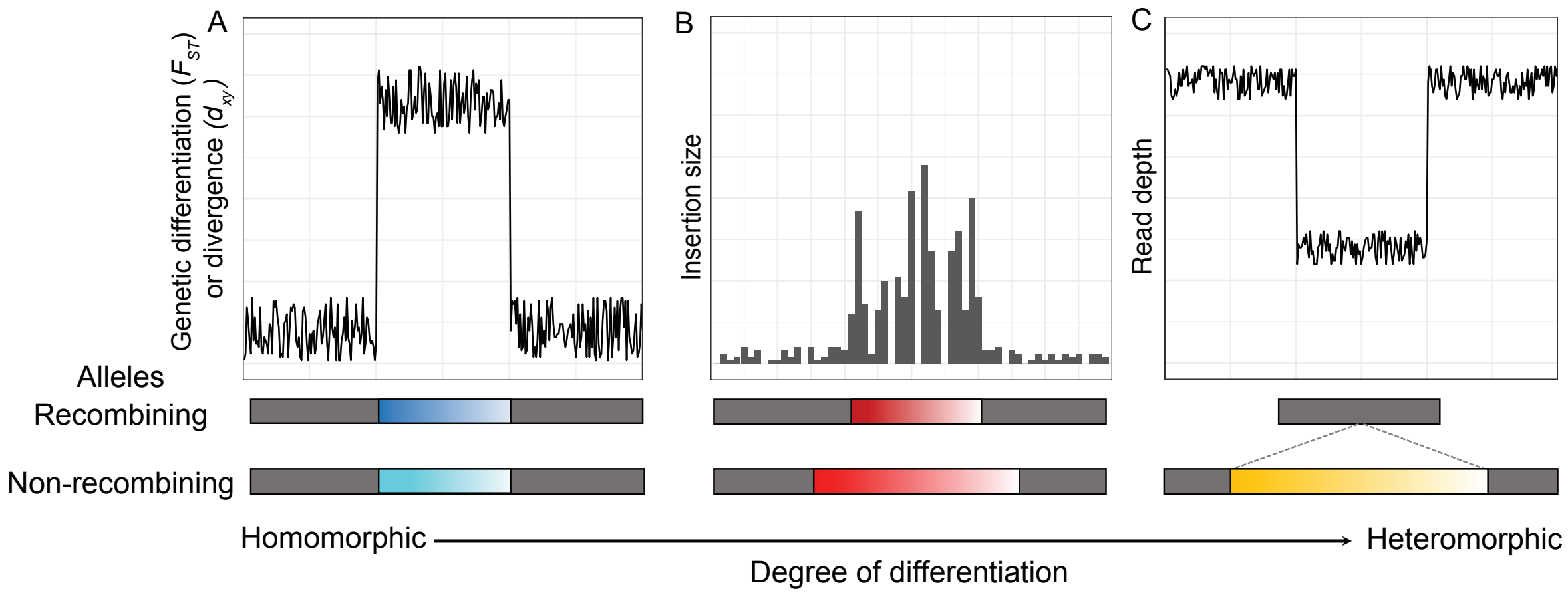

Figure Box 1. Signatures of divergence and degeneration as a result of suppressed recombination have been used to pinpoint the location of supergenes. Different strategies can be applied depending on the extent of differentiation between the recombining and non-recombining allele. A. Haplotypes remain homomorphic but mutations accumulate in the non-recombining one, such that genetic differentiation (e.g. $F_{ST}$) between morphs, or divergence ($d_{xy}$) between haplotypes can aid supergene identification (e.g. Tuttle et al. 2016). B. The non-recombining haplotype has expanded through the accumulation of repetitive elements, such that frequent and large insertions can indicate the occurrence of long-term suppressed recombination (e.g. Stolle et al. 2019). C. If there is hemizygosity at the supergene, analyses based on the detection of regions showing consistently reduced read depth relative to the rest of the genome in relevant morphs can aid in identification of the supergene (e.g. Li et al. 2016).

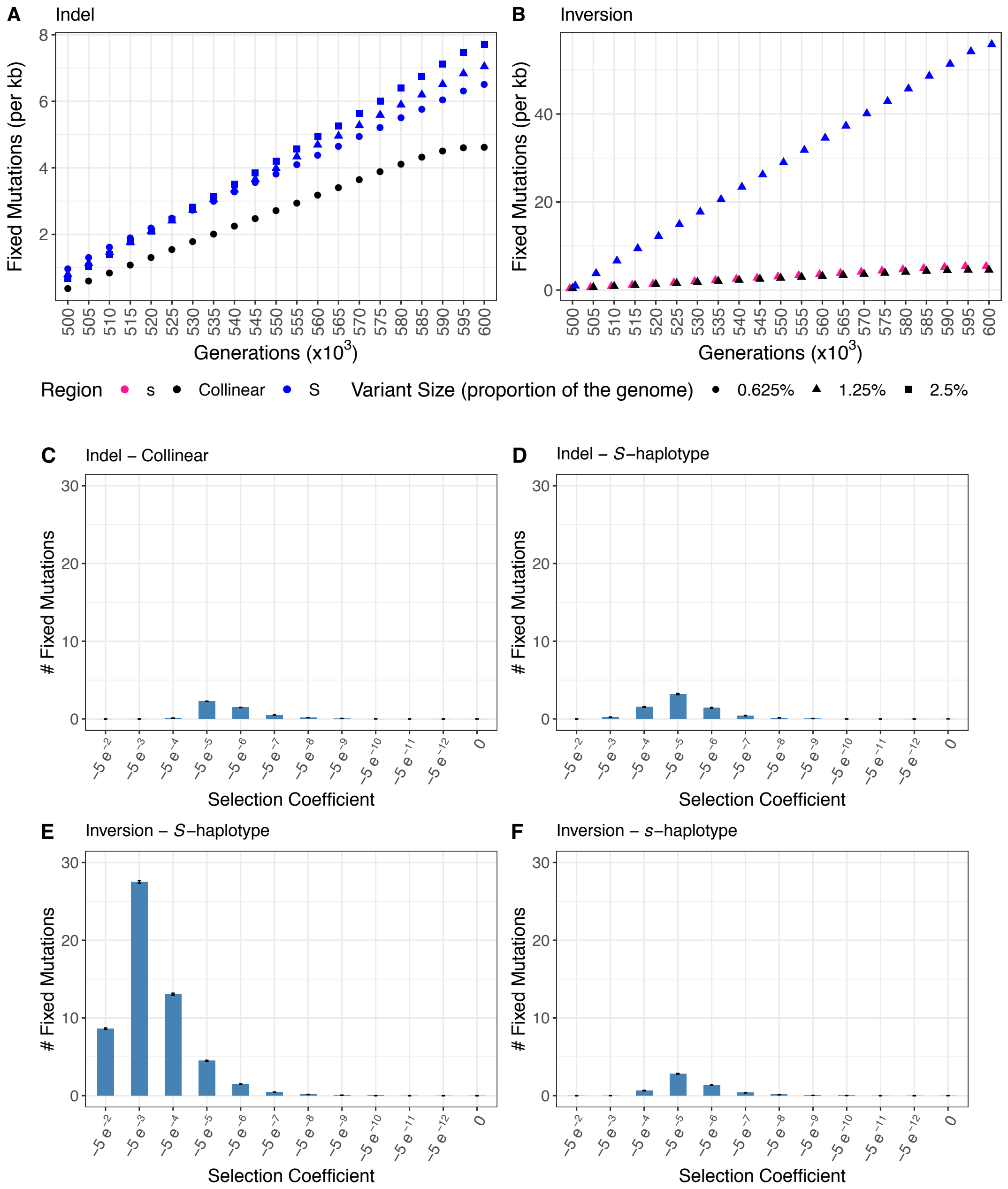

Figure Box 2. Dynamics of deleterious mutation accumulation in supergenes. A. Deleterious mutation accumulation over time is accelerated in the hemizygous *S*-haplotype (blue) compared to the collinear region (black). Shape indicates the size of the indel (circle - 0.625%, triangle 1.25%, square 2.5%). B. Deleterious mutation accumulation over time is accelerated in the inverted *S*-haplotype (blue) compared to the collinear region (black) and the standard arrangement *s*-haplotype (pink). C-F. Histogram showing selection coefficients of fixed mutations at the end of the simulation (600 k generations) in the (C) collinear region of the indel simulations, (D) hemizygous *S*-haplotype, (E) inversion *S*-haplotype, and (F) inversion standard arrangement *s*-haplotype. Error bars indicate ± standard error.

## Supplementary Methods

### Forward Population Genetic Simulations

We modeled an isolated population of diploid individuals at initial mutation-selection balance using SLiM v3.3.2 (Haller and Messer 2019). We simulated a population of $N$=50,000 diploid individuals. The genome consisted of two chromosomes of 5 Mb, 1 Mb of which were coding regions where allelic content was simulated. The allelic content of the rest of the chromosome was not simulated to alleviate the computational load, although recombination could occur anywhere. Coding regions, at which all sites could harbor selected mutations, were 100 kb segments, separated from each other by 400 kb of non-coding regions (i.e. areas where allelic content was not simulated).

We chose parameter estimates inspired by *Arabidopsis thaliana* to calibrate our model, because mutation and recombination rates are well characterized in this species. In our model, mutations happened at a rate of $\mu$=7 x $10^{-9}$ per bp per generation (Weng et al. 2019). All mutations only occurred in coding regions and affected individual fitness multiplicatively. The vast majority of mutations (99.9%) were deleterious and recessive with magnitudes of fitness effects ($|s|$) drawn from a Gamma distribution Γ (α=0.5, mean=0.0025). Beneficial mutations (0.1% of mutations) were co-dominant and drawn from an exponential distribution with a mean of 0.001. Overall recombination rate was set at 5.4 x $10^{-7}$ per base pair per meiosis following estimates from *A. thaliana* (Wijnker et al. 2013) but corrected for the high rate of selfing (all individuals in our simulations were obligate outcrossers). All simulations started following a burn-in of 500,000 generations to ensure that mutation-selection-drift equilibrium was attained.

*Modeling an Insertion*

We assumed that an insertion occurred in a random haplotype. The insertion occurred between two given loci on chromosome one and encompassed 0.625%, 1.25%, or 2.5% of the genome, corresponding to 31.25 kb, 62.5 kb or 125 kb prior to scaling (see below). For a given haplotype we modeled a 500-generation “invasion phase” and a subsequent 99,500-generation “evolution phase”. We performed 3 replicates of the “evolution phase” per haplotype. During the invasion phase, the insertion occurred in a random haplotype and was given a strong heterozygote advantage $s_{HET}$=0.0075 or $2Ns_{HET}$=750 to aid invasion. During invasion no recombination occurred in the inserted region, all mutations in the inserted region were dominant, and any insertion homozygote offspring were re-drawn.

After 500 generations of invasion the heterozygote advantage was removed and mating rules were added. We followed the mating rules of the S-locus in *Primula* only allowing for insertion heterozygotes (*S/s*) X standard homozygotes (*s/s*) matings. As in the invasion phase, no recombination occurred in the insertion heterozygotes (*S/s*), all mutations in the inserted region were dominant, and any insertion homozygote (*S/S*) offspring were re-drawn. The simulation ran for a total of 100,000 generations after the burn in. The final population was retained for calculating the distribution of fitness effects for various regions of the genome.

For each size of insertion we ran 3 replicates from 50 haplotypes.

*Modeling an Inversion*

Our indel model only allowed for crossing over which is suppressed within inverted regions. Thus, our inversion simulations allowed for gene conversion as well as crossing over. Recombination (i.e. double strand breaks) still occurred at the same rate but 47.5% of these because non-crossover gene conversion events. Gene conversion track length followed a Poisson distribution with parameter $\lambda = 400$ bp.

As before we assumed that the inversion occurred in a random haplotype. The inversion occurred between two given loci on chromosome one and encompassed 1.25% of the genome (i.e. 62.5 kb prior to rescaling, see below). The "invasion phase" and the "evolution phase" were identical to the insertion except that all mutations in the inverted region were recessive, and gene conversion could take place in inversion heterozygotes.

*Scaling*

Simulation with these parameters was not feasible because of the extremely large computational burden. Instead we used the common practice of rescaling parameters so that evolutionary processes happened at an accelerated rate (see for example Tuttle et al. 2016). We thus downscaled both population size and genome length by a factor 20 and upscaled the remaining parameters so that $2N\mu L$ and $2NrL$ (with $L$ the length of the genome) remained constant.

## Supplementary Results

### Invasion

Inversions invaded the population much more easily than insertions. This is likely due to mutation masking in the early stages of invasion for inversions when only heterokaryotypes occur. Invasion success for the hemizygous region, where all mutations are dominant, was lower and was strongly affected by the size of the region. This is because larger regions contained more deleterious mutations (Supplementary figure 1).

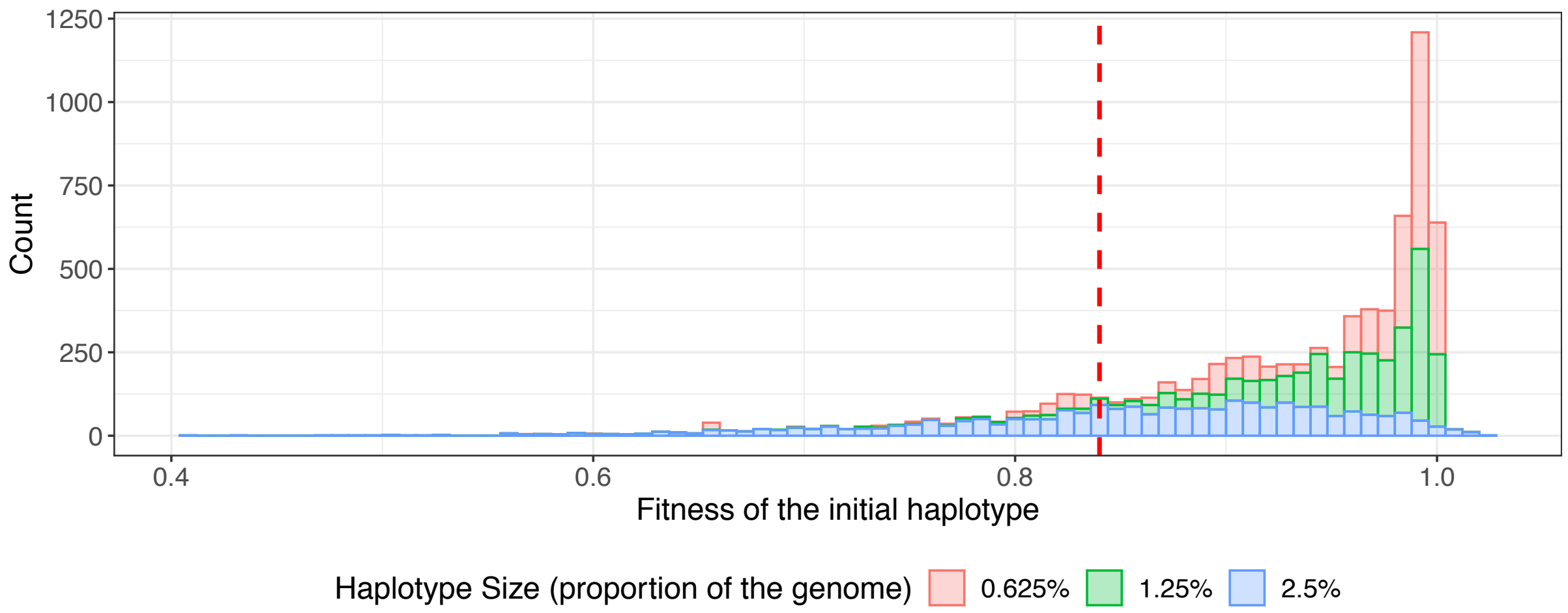

Supplemental Figure 1. Histogram depicting the fitness of initial S haplotype by size (red - 0.625% of the genome, green - 1.25%, blue - 2.5%). The dashed red line indicates the fitness needed for the haplotype to invade as an indel under our parameters.

## Literature Cited

Haller BC, Messer PW. 2019. SLiM 3: Forward genetic simulations beyond the Wright-Fisher model. Mol Biol Evol. 36:632-7.

Tuttle EM et al. 2016. Divergence and functional degradation of a sex chromosome-like supergene. Curr Biol. 26:344-350.

Weng ML et al. 2019. Fine-grained analysis of spontaneous mutation spectrum and frequency in *Arabidopsis thaliana*. Genetics. 211:703-14.

Wijnker E et al. 2013. The genomic landscape of meiotic crossovers and gene conversions in *Arabidopsis thaliana*. elife. 2:e01426.